\documentclass[%
 reprint,
 amsmath,amssymb,
 aps,
 pra, 
 longbibliography,
 floatfix
]{revtex4-2}

\usepackage{footmisc,multirow}
\usepackage{subfigure} 
\usepackage{amsmath,slashed}
\usepackage{bm}
\usepackage{upgreek}
\usepackage{graphicx}
\usepackage{color}
\usepackage{comment}

\begin{document}

\title{Multipole blackbody radiation shift of Rydberg energy levels}

\author{R. M. Potvliege}
\email{r.m.potvliege@durham.ac.uk}
\affiliation{Department of Physics, Durham University, Durham DH1 3LE, United Kingdom}



\begin{abstract}
We study the role of retardation in the energy shift of Rydberg states induced by thermal radiation, focusing on the case of temperatures higher than those for which the electric-dipole approximation is expected to apply. As anticipated by Farley and Wing [Phys. Rev. A {\bf 23}, 2397 (1981)], retardation needs to be taken into account in calculations of this energy shift around and above the temperature $\alpha\, mc^2/(3k_{\rm B}\,n^2)$, where $n$ is the principal quantum number of the state considered, $m$ is the
mass of the electron and $k_{\rm B}$ is Boltzmann constant --- e.g., around and above 1440~K for $n = 100$. The corresponding non-dipole shift dominates the electric-dipole shift at about 2.5 times that characteristic temperature. Thermal shifts arising from retardation and from the electric-quadrupole interaction are of the same order of magnitude as the diamagnetic thermal shift and would thus need to be taken into account in the circumstances where the latter is relevant.

\end{abstract}

\maketitle

\section{Introduction}
\label{section:intro}
 
Quantifying the AC Stark shift induced by the blackbody radiation (BBR) is often important in high precision spectroscopy and related areas. In particular, the requirements of modern atomic clocks may make it necessary to calculate the BBR shift beyond the electric-dipole approximation \cite{Porsev2006,Arora2012,Sahoo2014,Tang2024}. 
The current requirements of Rydberg states spectroscopy are less stringent. An estimate of the BBR shift based on the dipole approximation, thus neglecting retardation, normally suffices in this area.
How the BBR shift of Rydberg states varies within this approximation has been known for some time, in particular since the publication of a detailed study of the issue by Farley and Wing \cite{Farley1981} (some of the numerical results of Ref.~\cite{Farley1981} have recently been revisited \cite{Zalialiutdinov2022}).  In the present work, we focus on how this shift varies when this approximation breaks down, a question which does not seem to have been investigated in much detail so far.

The validity of neglecting retardation and multipole couplings when calculating the BBR shift of Rydberg states depends on the temperature and on the states. Farley and Wing \cite{Farley1981} noted that the dipole approximation should be expected to break down at temperatures such that the BBR field largely consists of spectral components varying significantly on the length scale of the state considered. This length scale can be taken to be the Bohr radius of the atom, i.e., $r_a = n_a^2$ for a state of principal quantum number $n_a$ (we use Hartree atomic units here and throughout this article, except where specified otherwise). Farley and Wing reckoned that retardation and multipole interactions may be expected to be significant at temperatures approaching or exceeding the state-dependent temperature $T_a$ at which $k_{\rm max}r_a = 1$, where $k_{\rm max}$ is the wavenumber at which the BBR spectral density is a maximum: 
\begin{equation}
T_a = \frac{c}{3\, n_a^2\,k_{\rm B}},
\label{eq:Ta}
\end{equation}
where $k_{\rm B}$ denotes the Boltzmann constant.
E.g., $T_a = 5770$~K for $n_a =50$, $T_a = 301$~K for $n_a = 219$ and $T_a = 160$~K for $n_a = 300$. This argument suggests that retardation significantly affects the BBR shift only at temperatures at which population redistribution by the BBR field \cite{Farley1981} and (possibly) BBR-induced ionization \cite{Beterov2009,Glukhov2022} are important.

All high Rydberg states have approximately the same BBR energy shift of $\pi (k_{\rm B}T)^2/(3\,c^3)$ within the dipole approximation \cite{Farley1981,Gallagher1979}. Namely, denoting the electric-dipole shift of state $a$ by $ \delta E_{a}^{({\rm E}1)}$,
\begin{equation}
    \delta E_{a}^{({\rm E}1)} \sim \frac{\pi}{3}\,\frac{(k_{\rm B}T)^2}{c^3}
\label{eq:asympt}
\end{equation}
for $n_a \rightarrow \infty$.
This asymptotic energy shift amounts to a frequency shift of 2417~Hz at $T = 300$~K. However, it has recently been pointed out that the BBR shift of Rydberg states also includes a contribution from the diamagnetic part of the interaction Hamiltonian and that this contribution may be significant for high enough principal quantum numbers and high enough temperatures \cite{Beloy2025,LopezRodriguez2025}. 
Ignoring retardation and relativistic corrections, the diamagnetic energy shift of a state $|a\rangle$ of a single active electron species is
\begin{equation}
\delta E_{a}^{({\rm D})} = \frac{\pi^3}{45}\,\frac{(k_{\rm B}T)^4}{c^5}\,
\langle a | r^2 | a \rangle,
\label{eq:dia}
\end{equation}
where $r$ denotes the distance of the active electron to the nucleus.
For a state of hydrogen with principal quantum number $n_a$ and orbital angular momentum quantum number $l_a$ \cite{Beloy2025,Bockasten1974},
\begin{equation}
\langle a | r^2 | a \rangle = \frac{1}{2}\, n_a^2[5n_a^2 -3l_a(l_a+1)+1].
\end{equation}
This energy shift translates into a frequency shift $\delta \nu_a^{({\rm D})}$ equal to $\delta E_{a}^{({\rm D})}/h$, where $h$ is the Planck constant. It amounts to about 1~Hz for the $50\,{\rm s}$ state of hydrogen at room temperature. While small for that state compared to the 2.4~kHz electric-dipole shift at the same temperature, $\delta E_{a}^{({\rm D})}$ increases rapidly with $n_a$ and may therefore contribute significantly to the energy difference between high Rydberg states, at least at sufficiently high temperatures \cite{Beloy2025}.

However, $\delta E_{a}^{({\rm D})}$ is not the only contribution of order $(k_{\rm B}T)^4/c^5$ to the BBR shift. We show that in good approximation, the BBR shift takes up the following form for a Rydberg state $|a\rangle$ (this equation does not apply for low-lying states):
\begin{align}
 \delta E_{a} \approx 
 \delta E_{a}^{({\rm E}1)} &+
 \frac{8\pi^3}{135}\,\frac{(k_{\rm B}T)^4}{c^5}\,
\langle a | r^2 | a \rangle \nonumber \\ &-\frac{44\pi^5}{42525}\,\frac{(k_{\rm B}T)^6}{c^7}\,
\langle a | r^4 | a \rangle + \cdots,
\label{eq:7}
\end{align}
The right-hand side is an asymptotic series which rapidly diverges when $T \agt T_a$ but can nonetheless be summed.
Its first term is the shift calculated within the dipole approximation. The next term is contributed by the diamagnetic shift $\delta E_{a}^{({\rm D})}$, by the electric-quadrupole term in the multipole expansion of the BBR shift and by a retardation correction. The higher order terms are contributed by higher electric-multipole shifts and other retardation corrections, including corrections to the diamagnetic shift.
Highly precise calculations of the BBR shift would require adding relativistic corrections of relative order $1/c^2$ to each of its terms as well as further corrections, also of relative order $1/c^2$, arising from the paramagnetic part of the interaction Hamiltonian. As explained  below, the error introduced by neglecting the relativistic corrections and the paramagnetic shift is negligible in the present context.

Eq.~(\ref{eq:7}) is obtained within a further approximation, which is to assume that the sum over intermediate states involved in the calculation of the electric-multipole shift of state $|a\rangle$ is dominated by the contributions of states close in energy to $|a\rangle$. We show, in Sec.~\ref{section:mainsection}, that this approximation is sound for the states and temperatures of interest here. Eq.~(\ref{eq:7}) can thus be used for characterizing the behavior of the BBR shift for Rydberg states and temperatures for which the dipole approximation is questionable. It can also be used for estimating the error incurred by approximating the BBR shift by the electric-dipole shift $\delta E_{a}^{({\rm E}1)}$.

Calculating the BBR shift with full allowance for retardation, as we do below, makes it possible to address the high temperature regime where $T$ approaches or exceeds the characteristic temperature $T_a$ of Eq.~(\ref{eq:Ta}). We show that the BBR shift becomes dominated by non-dipole corrections at $T \approx 2.5\, T_a$.

The rest of this article is organised in two main sections: we discuss the calculation of the electric-multipole shift without retardation in Sec.~\ref{section:mainsection}, and move to the formulation and calculation of the non-relativistic BBR shift with retardation in Sec.~\ref{section:retardation}. The main body of the paper ends with brief conclusions, in Sec.~\ref{secrtion:conclusions}. We only consider the case of an isotropic BBR field.
As already mentioned, we generally use Hartree atomic units, except where specified otherwise. We thus equate $c$ with the inverse of the fine structure constant, measure lengths in units of the Bohr radius, set $4 \pi \epsilon_0$ to 1 and express all energies in hartrees.

\section{BBR shift without retardation}
\label{section:mainsection}

\subsection{Non-relativistic theory}
\label{section:noretardation}

We consider a particular Rydberg state of the atom of interest, state $|a\rangle$ say, and the BBR energy shift of this state, $\delta E_{a}$. Ignoring retardation for the time being, we write $\delta E_{a}$ as a sum of electric-multipole shifts $\delta E_{a}^{({\rm E}K)}$, magnetic-multipole shifts $\delta E_{a}^{({\rm M}K)}$, and the diamagnetic shift defined by Eq.~(\ref{eq:dia}):
\begin{align}
    \delta E_{a}= \sum_{K = 1}^\infty \delta E_{a}^{({\rm E}K)} +  \sum_{K = 1}^\infty \delta E_{a}^{({\rm M}K)} + \delta E_{a}^{({\rm D})}.
    \label{eq:multipoleexp}
\end{align}
Following Ref.~\cite{Porsev2006},
\begin{align}
\delta E_{a}^{({\rm E}K)} &= -\frac{(k_{\rm B}T/c)^{2K+1}}{2j_a+1}\nonumber \\
&\qquad \times \sum_{p} |\langle a \Vert Q_K \Vert p\rangle|^2 
F_K\left(\frac{E_{p} - E_{a}}{k_{\rm B}T}\right),
\label{eq:deltaEKdefined}
\end{align}
where $\{|p\rangle\}$ is a complete set of eigenstates of the field-free Hamiltonian, $E_a$ is the energy of state $|a\rangle$, $E_p$ is the energy of state $|p\rangle$, ${T}$ is the temperature,
$Q_K$ is the electric-multipole operator of order $K$ and
\begin{align}
&F_K(y) = \frac{1}{\pi}\,\frac{K+1}{K(2K+1)!!(2K-1)!!}
\nonumber \\ & \qquad \times {\rm P.V.}\int_0^\infty\left(\frac{1}{y+x} + \frac{1}{y-x}\right)\frac{x^{2K+1}}{\exp(x)-1}\,{\rm d}x. 
\label{eq:FK}
\end{align}
The symbol ${\rm P.V.}$ indicates that the principal value of the integral is to be used in the calculation.
We will assume that each of the intermediate states $|p\rangle$ is characterized by a principal quantum number $n_p$ (or by a wave number $k_p$ for the continuum states), an orbital angular momentum quantum number $l_p$ and a total angular momentum quantum number $j_p$. Likewise, state $|a\rangle$ is characterized by the principal quantum number $n_a$, the orbital angular momentum quantum number $l_a$ and the total angular momentum quantum number $j_a$. Then, in terms of the radial matrix elements $\langle a | r^K | p \rangle$,
\begin{align}
&|\langle a \Vert Q_K \Vert p\rangle|^2 = 
(2j_a+1)(2j_p+1)(2l_a+1)(2l_p+1) \nonumber \\
& \qquad \qquad \times 
\begin{Bmatrix} l_a & l_p & K \\ j_p & j_a & 1/2 \end{Bmatrix}^2
\begin{pmatrix}    l_a & K & l_p \\ 0 & 0 & 0 \end{pmatrix}^2 \left| \langle a | r^K | p \rangle\right|^2.
\end{align}
We also assume that a single-active-electron model of the atom is sufficiently accurate for the states of interest, in which case
\begin{equation}
\langle a | r^K | p \rangle = \int_0^\infty R_a^{*}(r)  \, r^K \, R_p(r)\,r^2\,{\rm d}r,
\label{eq:matel}
\end{equation}
where $R_a(r)$ and $R_p(r)$ are the radial wave functions of the respective states.
The summation over $j_p$ implied by Eq.~(\ref{eq:deltaEKdefined}) is straightforward in the approximation where $E_p$ and $R_p(r)$ are the same for all the components of a same fine structure manifold, since \cite{Edmonds}
\begin{equation}
\sum_{j_p} (2j_p+1) (2l_a + 1) \begin{Bmatrix} l_a & l_p & K \\ j_p & j_a & 1/2 \end{Bmatrix}^2 = 1.
\end{equation}
In this non-relativistic approximation,
\begin{align}
&\delta E_{a}^{({\rm E}K)} = -\left(\frac{k_{\rm B}T}{c}\right)^{2K+1} \sum_{n_pl_p} (2l_p+1) \, \times \nonumber \\
&\quad \begin{pmatrix}    l_a & K & l_p \\ 0 & 0 & 0 \end{pmatrix}^2 
|\langle n_al_a | r^K| n_pl_p\rangle|^2
F_K\left(\frac{E_{p} - E_{a}}{k_{\rm B}T}\right).
\label{eq:deltaEKdefinednr}
\end{align}
The electric-dipole shift $\delta E_{a}^{({\rm E}1)}$ has been studied by Farley and Wing \cite{Farley1981}. We discuss the electric-quadrupole shift in Secs.~\ref{section:Hydrogen} and \ref{section:Cesium} and the higher electric-multipole shifts in Sec.~\ref{section:higher}. We revisit these results in Sec.~\ref{section:retardation}, then with retardation taken into account.

The function $F_K(y)$ takes on a much simpler form when $|y| \ll 1$: for $|y| \rightarrow 0$,
\begin{equation}
F_K(y) \sim -\frac{2y}{\pi}\,\frac{(K+1)(2K-1)!}{K(2K+1)!!(2K-1)!!}\,\zeta(2K)
\label{eq:FKsmally}
\end{equation}
where $\zeta(\cdot)$ is the Riemann zeta function \cite{zeta}. Of particular interest in the following are the energy shifts $\delta {E}_{a}^{({\rm E}K)\prime}$, which are defined in the same way as the shifts $\delta E_{a}^{({\rm E}K)}$ but with the function $F_K(y)$ replaced by its small-$|y|$ form:
\begin{align}
\delta E_{a}^{({\rm E}K)\prime} &= \frac{1}{\pi} \frac{(K+1)(2K-1)!\,\zeta(2K)}{K(2K+1)!!(2K-1)!!}\, \frac
{(k_{\rm B}T)^{2K}}{c^{2K+1}}\,\times \nonumber \\ & \qquad \quad
\frac{2}{2j_a+1}\,
\sum_{p} (E_p - E_a) |\langle a \Vert Q_K \Vert p\rangle|^2.
\label{eq:deltaEKapprox}
\end{align}
In the non-relativistic approximation of Eq.~(\ref{eq:deltaEKdefinednr}),
\begin{align}
&\delta E_{a}^{({\rm E}K)\prime} = \frac{2}{\pi} \frac{(K+1)(2K-1)!\,\zeta(2K)}{K(2K+1)!!(2K-1)!!}\, \frac
{(k_{\rm B}T)^{2K}}{c^{2K+1}}\,\times \nonumber \\ & \;
\sum_{n_pl_p} (2l_p+1) \begin{pmatrix}    l_a & K & l_p \\ 0 & 0 & 0 \end{pmatrix}^2 (E_p-E_a) 
\left|\langle n_al_a | r^K| n_pl_p\rangle\right|^2.
\label{eq:deltaEKapprox2}
\end{align}
The relevance of the shifts $\delta E_{a}^{({\rm E}K)\prime}$ stems from the fact that for Rydberg states the shifts $\delta E_{a}^{({\rm E}K)}$ are
dominated by the contribution of intermediate states  
for which $|(E_p-E_a)/k_{\rm B}T| \ll 1$. This fact is noted in Ref.~\cite{Farley1981} in regard to the electric-dipole shift and is established for the higher electric-multipole shifts in Sec.~\ref{section:quadrupole}. Thus $\delta E_{a}^{({\rm E}K)} \approx \delta E_{a}^{({\rm E}K)\prime}$ for Rydberg states. 

Evaluating the shifts $\delta E_{a}^{({\rm E}K)\prime}$ is particularly simple as the summation over intermediate states can be done analytically. Indeed, as shown in Appendix~\ref{section:appendix},
\begin{align}
&\sum_{n_pl_p} (2l_p+1) \begin{pmatrix}    l_a & K & l_p \\ 0 & 0 & 0 \end{pmatrix}^2 (E_p-E_a) 
\left|\langle n_al_a | r^K| n_pl_p\rangle\right|^2 \nonumber \\
&\qquad \qquad \qquad \qquad = \frac{1}{2}\, K(2K+1)\, \langle a | r^{2K-2} | a \rangle
\label{eq:deltaEKapprox2bis}
\end{align}
for hydrogenic species. Thus
\begin{align}
\delta {E}_{a}^{({\rm E}K)\prime} &= \frac{1}{\pi} \frac{(K+1)(2K+1)!\,\zeta(2K)}{2K(2K+1)!!(2K-1)!!} \frac
{(k_{\rm B}T)^{2K}}{c^{2K+1}}\,\times \nonumber \\ & \qquad \qquad \qquad \qquad \qquad \qquad  \quad
\langle a | r^{2K-2} | a \rangle
\label{eq:EKfinal2}
\end{align}
for hydrogen and one-electron ions.
In particular,
\begin{equation}
\delta E_{a}^{({\rm E}1)\prime} = \frac{\pi}{3}\,\frac{(k_{\rm B}T)^2}{c^3}
\label{eq:E1tilde}
\end{equation}
for the electric-dipole shift and
\begin{equation}
\delta E_{a}^{({\rm E}2)\prime} = \frac{\pi^3 (k_{\rm B}T)^4}{45\,c^5}\,
\langle a | r^2 | a \rangle
\label{eq:E2andD}
\end{equation}
for the electric-quadrupole shift.
The numerical results presented in Sec.~\ref{section:Cesium} indicate that Eq.~(\ref{eq:EKfinal2}) also holds for Rydberg states of other systems, at least approximately.
Eq.~(\ref{eq:E1tilde}) applies generally \cite{Farley1981}.

It can be noted, by comparing Eq.~(\ref{eq:E2andD}) to Eq.~(\ref{eq:dia}), that $\delta E_{a}^{({\rm E}2)\prime} = \delta E_{a}^{({\rm D})}$. However, it is shown in Sec.~\ref{section:Eret} that $\delta E_{a}^{({\rm E}2)\prime}$ is largely cancelled by a term proportional to $(k_{\rm B}T)^4
\langle a | r^2 | a \rangle/c^5$ contributed by retardation.

The magnetic-multipole shifts $\delta E_a^{({\rm M}K)}$ are smaller by a factor $1/c^2$ compared to their electric-multipole counterparts. They are considered in Sec.~\ref{section:Pret}.

\subsection{Numerical illustration}
\label{section:quadrupole}

\subsubsection{Methods}
\label{section:methods}

In the case of hydrogen, we calculate the electric-dipole shift $\delta E_{a}^{({\rm E}1)}$ and electric-quadrupole shift $\delta E_{a}^{({\rm E}2)}$ without retardation using the method outlined in Appendix~D of Ref.~\cite{Jones2020}. That is, we construct the matrix ${\sf H}$ representing the non-relativistic Hamiltonian in a basis of Sturmian functions and spherical harmonics and solve the corresponding generalized eigenvalues problem, ${\sf H}\,{\sf c}_k = w_k {\sf S}\,{\sf c}_k$, where ${\sf S}$ is the overlap matrix of the basis. The eigenvalues $w_k$ effectively form a discrete representation of the whole spectrum of the atom, including its continuum part. We use 300 to 600 radial Sturmians for each value of $l$, which ensures an excellent agreement of all the 70 lowest values $w_k$ with the corresponding eigenenergies of the exact Hamiltonian, $-1/2n_k^2$ with $n_k = l+1$, $l+2$,\ldots~Doing so makes it possible to replace the summation over the intermediate states $|p\rangle$, which involves an integration over continuum states, by a simpler summation over the generalized eigenvectors of the matrix ${\sf H}$. The basis used in the calculation is large enough to ensure an excellent agreement between the resulting values of the matrix elements  $\langle a | r | p \rangle$ and $\langle a | r^2 | p \rangle$ and their exact values \cite{Sanchez1993} for all bound states of interest, an excellent agreement of the static dipole and quadrupole polarizabilities of these states with their analytical values \cite{Mei2020},
and an excellent agreement between the values of $\delta E^{({\rm E}1)\prime}$ and $\delta E^{({\rm E}2)\prime}$ calculated as per Eq.~(\ref{eq:deltaEKapprox2}) and those calculated as per Eqs.~(\ref{eq:E1tilde}) and (\ref{eq:E2andD}).
The method will be described in greater detail elsewhere \cite{Potvliege2026}.

This approach does not easily generalize to other atoms. We used the {\sc ARC} program \cite{ARCpapers,ARCprogs} to compute the state energies and quadrupole matrix elements required by the calculation of the BBR quadrupole shift in cesium reported in Sec.~\ref{section:Cesium}.  

Computing the shifts $\delta E_a^{({\rm E1})}$ and $\delta E_a^{({\rm E2})}$ also involves the calculation of the $F_K$ function for $K=1$ and 2.
As in Ref.~\cite{Jones2020}, we do this by contour integration in the complex plane.

For convenience, we state the resulting energy shifts in terms of the corresponding frequency shifts, and similarly for the energy shifts considered later in this paper: 
\begin{equation}
\delta \nu_a^{(\cdot)} \equiv \delta E_a^{(\cdot)}/h, \qquad
\delta \nu_a^{(\cdot)\prime} \equiv \delta E_a^{(\cdot)\prime}/h,
\end{equation}
where $h$ is the Planck constant.

\subsubsection{The electric-dipole shift in hydrogen}
\label{section:Hydrogendipole}
\begin{figure}[t!]
\centering
\includegraphics[width=\columnwidth]{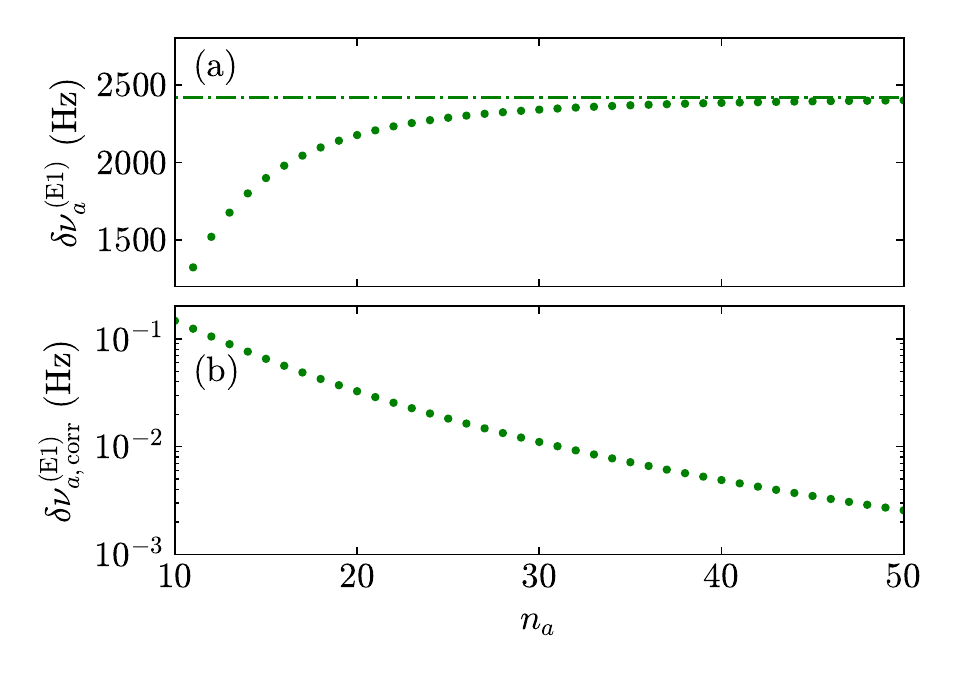}
\caption{
(a) The electric-dipole frequency shifts $\delta {\nu}_a^{({\rm E}1)}$ (markers) and $\delta \nu_a^{({\rm E}1)\prime}$ (dash-dotted line) of s-states of hydrogen at 300~K vs.\ the principal quantum number of the state. (b) The corresponding relativistic correction to the electric-dipole shift as defined by Eq.~(\ref{eq:relcorr}).  
}
\label{fig:Fig_rel_vs_nrel}
\end{figure}
For completeness, the variation of the electric-dipole energy shifts $\delta {E}_a^{({\rm E}1)}$ with the principal quantum number of the state is illustrated by Fig.~\ref{fig:Fig_rel_vs_nrel}(a) for the case of s-states of hydrogen at room temperature. (Part (b) of this figure is discussed in Sec.~\ref{section:relatcorr}.) As noted above and in Ref.~\cite{Farley1981}, $\delta {E}_a^{({\rm E}1)}$ is well approximated by the $n_a$-independent shift of Eq.~(\ref{eq:E1tilde}) for sufficiently high Rydberg states.

\subsubsection{The electric-quadrupole shift in hydrogen}
\label{section:Hydrogen}

\begin{figure}[t!]
\centering
\includegraphics[width=\columnwidth]{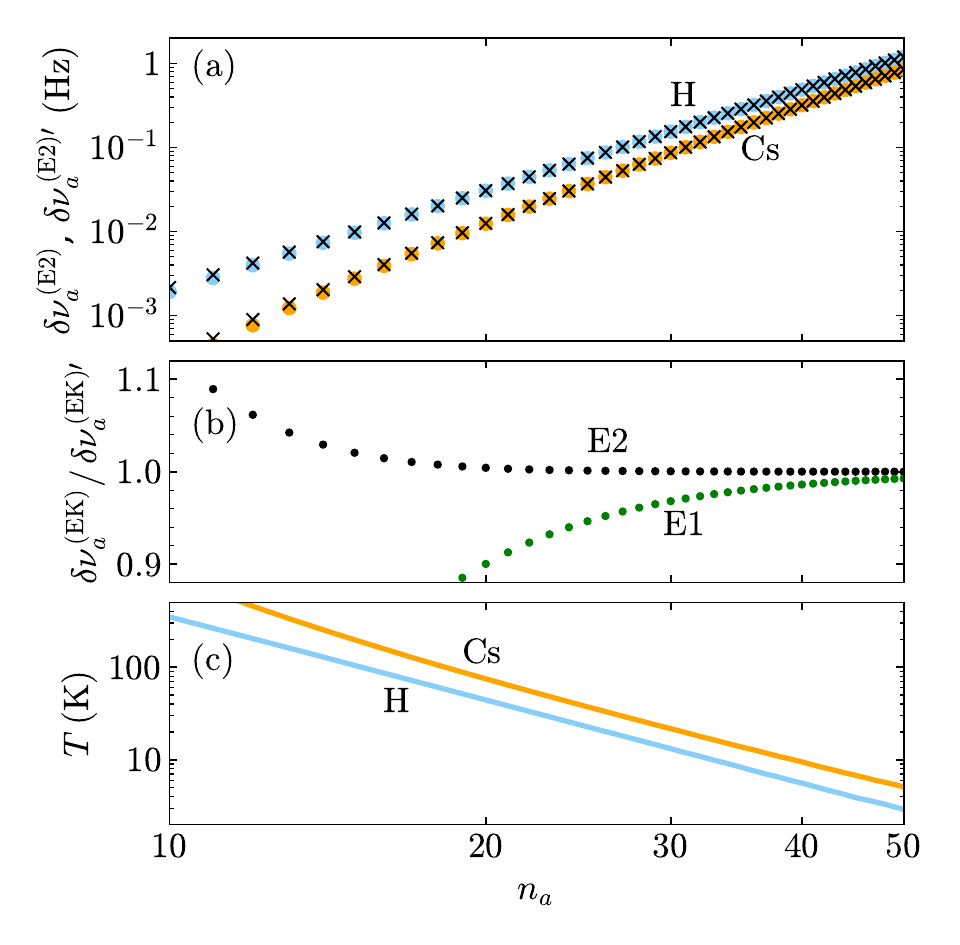}
\caption{
(a) The electric-quadrupole BBR shifts $\delta {\nu}_a^{({\rm E}2)}$ (crosses) and $\delta \nu_a^{({\rm E}2)\prime}$ (filled circles) of s-states of hydrogen and cesium at 300~K vs.\ the principal quantum number of the state. (b) The ratios $\delta {\nu}_a^{({\rm E}2)}/\,{\delta} \nu_a^{({\rm E}2)\prime}$ (black markers) and $\delta {\nu}_a^{({\rm E}1)}/\,{\delta} \nu_a^{({\rm E}1)\prime}$ (green markers) for the s-states of hydrogen at 300~K. (c) The temperature below which ${\delta} \nu_a^{({\rm E}2)\prime}$ differs from $\delta \nu_a^{({\rm E}2)}$ by more than 10\%.
}
\label{fig:Fig1}
\end{figure}
How the BBR frequency shifts $\delta \nu_a^{({\rm E}2)}$ and $\delta \nu_a^{({\rm E}2)\prime}$ vary with principal quantum number in the case of hydrogen is shown in parts (a) and (b) of Fig.~\ref{fig:Fig1} ($T=300$~K, as in Fig.~\ref{fig:Fig_rel_vs_nrel}). Only results for the s-series are presented in Fig.~\ref{fig:Fig1}, for clarity. The results for the p- and d-series are almost identical \cite{Potvliege2026}.

While very small for the low lying states, $\delta \nu_a^{({\rm E}2)}$ increases rapidly with $n_a$, roughly like $n_a^4$, to reach values of the order of 1~Hz for $n_a \approx 50$ at the temperature considered \cite{77K}. Moreover, $\delta \nu_a^{({\rm E}2)}$ and $\delta \nu_a^{({\rm E}2)\prime}$ are almost equal for large values of $n_a$: these two shifts differ by less than 1\% for $n_a \geq 20$ and the difference decreases as $n_a$ increases. The rapid convergence of $\delta \nu_a^{({\rm E}2)}$ to $\delta \nu_a^{({\rm E}2)\prime}$ for high Rydberg states can be noted from Fig.~\ref{fig:Fig1}(b). This figure also shows that $\delta \nu_a^{({\rm E}2)}$ converges to $\delta \nu_a^{({\rm E}2)\prime}$ faster than $\delta \nu_a^{({\rm E}1)}$ converges to $\delta \nu_a^{({\rm E}1)\prime}$.

\renewcommand{\arraystretch}{1.2}
\begin{table}
\caption{Electric-quadrupole BBR shift of the $50\,{\rm s}$ state of hydrogen at a temperature of 300~K as calculated in various approximations. Retardation is neglected. $\delta \nu_a^{({\rm E2})}$: results obtained when the $F_2(y)$ function is calculated according to its definition, Eq.~(\ref{eq:FK}). $\delta \nu_a^{({\rm E2})\prime}$: results obtained when the $F_2(y)$ function is approximated by its small-$|y|$ limit, Eq.~(\ref{eq:FKsmally}). The first five rows give the results obtained when the summation over the principal quantum number of the intermediate state is restricted to the range indicated. The penultimate row are the results obtained by summing over a complete set of intermediate states.}
\label{table:Table1}
\begin{center}
\begin{ruledtabular}
\begin{tabular}{ccc}
Range & $\delta \nu_a^{({\rm E2})}$ (Hz) & $\delta \nu_a^{({\rm E2})\prime}$ (Hz) \\[1mm]
\colrule
$48 \leq n_p \leq 52$
& 1.13113 & 1.13112 \\
$46 \leq n_p \leq 54$
& 1.17300 & 1.17299 \\
$44 \leq n_p \leq 56$
& 1.18321 & 1.18320  \\
$42 \leq n_p \leq 58$
& 1.18732 & 1.18731\\
$40 \leq n_p \leq 60$
& 1.18941 & 1.18939\\
Limit &    1.19428 & 1.19425 \\
Eq.~(\ref{eq:E2andD}) & & {1.19425}
\end{tabular}
\end{ruledtabular}
\end{center}
\end{table}
\renewcommand{\arraystretch}{1.2}
\begin{table}
\caption{The same as Table~\ref{table:Table1} but here for the electric-quadrupole BBR shift of the $(n=50,l=49)$ circular state.}
\label{table:Table1circ}
\begin{center}
\begin{ruledtabular}
\begin{tabular}{ccc}
Range & $\delta \nu_a^{({\rm E2})}$ (Hz) & $\delta \nu_a^{({\rm E2})\prime}$ (Hz) \\[1mm]
\colrule
$48 \leq n_p \leq 52$
& 0.26698 & 0.26699 \\
$48 \leq n_p \leq 54$
& 0.48887 & 0.48884 \\
$48 \leq n_p \leq 56$
& 0.49197 & 0.49194  \\
$48 \leq n_p \leq 58$
& 0.49210 & 0.49208\\
$48 \leq n_p \leq 60$
& 0.49211 & 0.49209\\
Eq.~(\ref{eq:E2andD}) & & {0.49209}
\end{tabular}
\end{ruledtabular}
\end{center}
\end{table}
The similarity between the values of $\delta \nu_a^{({\rm E}2)}$ and of $\delta \nu_a^{({\rm E}2)\prime}$ indicates that replacing $F_2(y)$ by its small-$|y|$ limit is a good approximation for the intermediate states contributing most to $\delta \nu_a^{({\rm E}2)}$, at least at 300~K. That this is indeed the case for high Rydberg states is shown by the 5-figure agreement between the values of  $\delta \nu_a^{({\rm E}2)}$ and $\delta \nu_a^{({\rm E}2)\prime}$ given in the penultimate row of Table~\ref{table:Table1}.
These two results were obtained by summing over all intermediate states, including those representing the discretized continuum, and are converged with respect to this summation to the number of digits given in the table. The converged value of $\delta \nu_a^{({\rm E}2)\prime}$ is itself in full agreement with the result predicted by Eq.~(\ref{eq:E2andD}).

How these converged values build up when summing over the intermediate states is illustrated by the first five rows of the table, which give the shifts as calculated when this summation is restricted to the range of principal quantum numbers indicated. The energy shift of these large $n_a$-states is largely contributed by intermediate states with $n_p \approx n_a$. The argument of the $F_2(y)$ function is close to 0 for such intermediate states. E.g., at 300~K, $|(E_p-E_a)/k_{\rm B}T|$ is at most 0.12 for $40 \leq n_p \leq 60$ when $n_a = 50$. The agreement between $\delta \nu_a^{({\rm E}2)}$ and $\delta \nu_a^{({\rm E}2)\prime}$ shows that the intermediate states with a higher energy difference do not contribute enough for invalidating the replacement of $F_2(y)$ by its small-$|y|$ form. The sum over intermediate state converges rapidly for lower states, too: when $n_a \leq 10$, and for $T= 300$~K, the value of $\delta \nu_a^{({\rm E}2)}$ calculated with $n_p$ limited to the range $[n_a-10,n_a+10]$ never differs from the converged value by more than 0.5\%. Convergence is also found for states of high orbital angular momentum quantum numbers such as circular Rydberg states (Table~\ref{table:Table1circ}).

Since the argument of the $F_2$ function is inversely proportional to $k_{\rm B}T$, the approximation made by replacing $F_2(y)$ by its small-$y$ form deteriorates when $T$ decreases. As can be seen from Fig.~\ref{fig:Fig1}(c), it remains nonetheless good down to temperatures close to 3~K for $n_a \agt 50$.

\subsubsection{The electric-quadrupole shift in cesium}
\label{section:Cesium}

The small importance of the intermediate states outside the $[n_a-10,n_a+10]$ range noted above for the quadrupole shift of hydrogen suggests that these states may also be neglected, in good approximation, when calculating the quadrupole BBR shift of the Rydberg states of alkali metals. Doing so for cesium yields the other set of results shown in Fig.~\ref{fig:Fig1}(a), for which the $\delta \nu_a^{({\rm E}2)\prime}$ shifts are now represented by orange markers rather than by blue markers. 

The results are qualitatively the same as for hydrogen, although the conditions of applicability of the proof of Eq.~(\ref{eq:deltaEKapprox2bis}) given in Appendix~\ref{section:appendix} are not strictly met here (we take the fine structure splitting into account and {\sc ARC} uses different model potentials for different series when calculating the radial wave functions, while the proof assumes no fine structure splitting and the same potential for all relevant series).  

\subsubsection{Higher multipole shifts}
\label{section:higher}

\renewcommand{\arraystretch}{1.2}
\begin{table}
\caption{$K=8$ electric-multipole BBR shift of the $50\,{\rm s}$ state of hydrogen at a temperature of 300~K as calculated in various approximations. Retardation is neglected. $\delta \nu_a^{({\rm E8})}$: results obtained when the $F_8(y)$ function is calculated according to its definition, Eq.~(\ref{eq:FK}). $\delta \nu_a^{({\rm E}8)\prime}$: results obtained when the $F_8(y)$ function is approximated by its small-$|y|$ limit, Eq.~(\ref{eq:FKsmally}). As in Table~\ref{table:Table1}, the first five rows give the results obtained when the summation over the principal quantum number of the intermediate state is restricted to the range indicated.}
\label{table:Table2}
\begin{center}
\begin{ruledtabular}
\begin{tabular}{ccc}
Range & $\delta \nu_a^{({\rm E8})}$ ($10^{-18}$~Hz) & $\delta \nu_a^{({\rm E8})\prime}$ ($10^{-18}$~Hz)\\[1mm]
\colrule
$48 \leq n_p \leq 52$
& 2.25360 & 2.25360 \\
$46 \leq n_p \leq 54$
& 2.25424 & 2.25424 \\
$44 \leq n_p \leq 56$
& 2.25433 & 2.25432   \\
$42 \leq n_p \leq 58$
& 2.25433 & 2.25433\\
$40 \leq n_p \leq 60$
& 2.25433  & 2.25433 \\
Eq.~(\ref{eq:EKfinal2}) & & {2.25433}
\end{tabular}
\end{ruledtabular}
\end{center}
\end{table}

We have also evaluated the higher order multipole contributions to the BBR shift of hydrogen up to $K=8$, here by summing the contributions of bounds intermediate states energetically close to the state of interest rather than by summing the contributions of a large number of Sturmian basis functions \cite{Sanchez1993}. We found that
the higher multipole BBR shifts of Rydberg states are also dominated by the contributions of neighboring intermediate states for which $|E_p - E_a|/k_{\rm B}T \ll 1$. For instance, including only intermediate states with $n_a - 10 \leq n_p \leq n_a+10$ in the sum is sufficient for obtaining values of the shifts $\delta E^{({\rm E}K)\prime}$ verifying Eq.~(\ref{eq:EKfinal2}) to six significant digits for $n_a \approx 50$. In fact, the higher $K$ is, the more dominant the neighboring states are --- e.g., compare the results given in Table~\ref{table:Table1} for the quadrupole shift with those given in Table~\ref{table:Table2} for the $K=8$ shift. As a result, the multipole shifts $\delta {E}^{({\rm E}K)}$ are increasingly well approximated by $\delta E^{({\rm E}K)\prime}$, thus by the right-hand side of Eq.~(\ref{eq:EKfinal2}): for all practical purposes,
\begin{align}
&\delta {E}_{a}^{({\rm E}K)} = \frac{1}{\pi} \frac{2^{2K-1}(K-1)!(K+1)!}{(2K)!}\, \zeta(2K)\nonumber \\ &  \qquad \times \frac
{(k_{\rm B}T)^{2K}}{c^{2K+1}}\, 
\langle a | r^{2K-2} | a \rangle, \quad K \geq 2,\, n_a \gg 10.
\label{eq:EKfinal3}
\end{align}
This approximation improves not only when $K$ increases for a fixed temperature but also when $T$ increases for a fixed $K$ (these high multipole shifts increase rapidly with $T$ and are numerically significant for $T \approx T_a$).

\subsection{Relativistic corrections}
\label{section:relatcorr}

We estimate the importance of the relativistic corrections to $\delta E_a^{({\rm E}1)}$ and $\delta E_a^{({\rm E}2)}$ in hydrogen by repeating the above calculations with the non-relativistic Hamiltonian, $H$, replaced by the  Breit-Pauli Hamiltonian, ${H}_{\rm BP}$:
\begin{equation}
    {H}_{\rm BP} = {H}_{} - \frac{{\bf p}^4}{8c^2}+\frac{1}{2c^2}\frac{{\bf L}\cdot{\bf S}}{r^3} +
    \frac{\pi}{2c^2}\delta({\bf r}),
\end{equation}
where ${\bf p}$, ${\bf L}$ and ${\bf S}$ are the momentum operator, orbital angular momentum operator and spin operator of the electron. The Breit-Pauli Hamiltonian is consistent with the Dirac Hamiltonian to order $1/c^2$. Diagonalizing it on the Sturmian basis and using the resulting wave functions and energies in Eq.~(\ref{eq:deltaEKdefined}) yields electric-dipole shifts $\delta E^{({\rm E}1)}_{a,{\rm BP}}$ and electric-quadrupole shifts $\delta E^{({\rm E}2)}_{a,{\rm BP}}$. These shifts differ from their non-relativistic counterparts by terms of order $1/c^2$ as well as by terms of higher order in $1/c^2$. The latter are not consistent with a formulation based on the Breit-Pauli Hamiltonian. However, they are negligible for the states considered. We can therefore quantify the importance of the relativistic corrections to the electric-dipole shift by the difference $\delta E^{({\rm E}1)}_{a,{\rm BP}} -\delta E^{({\rm E}1)}_{a}$, or equivalently by the corresponding frequency shift $\delta \nu^{({\rm E}1)}_{a,{\rm corr}}$:
\begin{equation}
\delta \nu^{({\rm E}1)}_{a,{\rm corr}} = \left[\delta E^{({\rm E}1)}_{a,{\rm BP}} -\delta E^{({\rm E}1)}_{a} \right]/h,
\label{eq:relcorr}
\end{equation}
where $\delta E^{({\rm E}1)}_{a}$ is the shift calculated non-relativistically.
The small importance of the relativistic corrections for the Rydberg states and temperatures considered in the present work is borne out by Fig.~\ref{fig:Fig_rel_vs_nrel}(b). E.g., at 300~K, $\delta \nu^{({\rm E}1)}_{a,{\rm corr}} \approx 3$~mHz for the 50~s state. By contrast, $\delta \nu^{({\rm E}2)}_a$ and the leading retardation correction to $\delta \nu^{({\rm E}2)}_a$ are both of the order of $1$~Hz for that state at this temperature. We will thus neglect these relativistic corrections in the following. However, we note that these corrections may exceed the quadrupole and diamagnetic BBR shifts for much lower excited states and/or for cryogenic temperatures.

The relativistic corrections to the electric-quadrupole shift $\delta E_a^{({\rm E}2)}$ are also negligible for our purposes --- e.g.,
$\delta E_{a,{\rm BP}}^{({\rm E}2)}$ differs from $\delta E_a^{({\rm E}2)}$ by not more than 0.003\% for any s-state with $n_a \geq 10$ at room temperature.

\section{BBR shift with retardation}
\label{section:retardation}

\subsection{Non-relativistic theory}
\label{section:theory}

We start by considering the AC Stark shift of a state $|a\rangle$ induced by an electromagnetic field with wave vector ${\bf k}$ and polarization vector $\bm{\hat \epsilon}_{}$ ($|{\bf k}| = k = \omega/c$, ${\bf k}\cdot \bm{\hat \epsilon_{}} = 0$ and $\bm{\hat \epsilon_{}}\cdot \bm{\hat \epsilon_{}}^* = 1$). Following Ref.~\cite{Beloy2025}, we define its electric field and magnetic field components by the equations
\begin{subequations}
\begin{align}
\label{eq:Ek}
{\bf E}_{{\bf k}}({\bf r},t) &= \frac{{\cal E}}{2}\,
\left(\bm{\hat \epsilon}_{}\exp[i({\bf k}\cdot{\bf r}-\omega t) ] + \mbox{c.c.}\right),\\
\label{eq:Bk}
{\bf B}_{{\bf k}}({\bf r},t) &= \frac{{\cal E}}{2c}\,
\left(\bm{\hat \beta}_{}\exp[i({\bf k}\cdot{\bf r}-\omega t) ] + \mbox{c.c.}\right),
\end{align}
\end{subequations}
with
${\bm {\hat \beta}}_{}={\bf k}{\bm \times} {\bm{\hat \epsilon}}_{}/k$.
The Hamiltonian of a one-electron system in the presence of this field takes up the following form in the Power-Zienau-Woolley gauge \cite{Loudon,Komninos2002,Anzaki2018,noteaboutPZW}:
\begin{align}
H &= \frac{1}{2}\left[-i\nabla -\int_0^1 {\bf r}{\bm \times} {\bf B}_{{\bf k}}(\lambda {\bf r},t)\,\lambda\, {\rm d}\lambda \right]^2 + V(r) \nonumber \\
&  \qquad \qquad \qquad \qquad \qquad + \int_0^1 {\bf r}\cdot{\bf E}_{{\bf k}}(\lambda {\bf r},t)\,{\rm d}\lambda.
\label{eq:HPZW}
\end{align}
To second order in ${\cal E}$, 
the electric and magnetic fields ${\bf E}_{{\bf k}}({\bf r},t)$ and ${\bf B}_{{\bf k}}({\bf r},t)$ shift
the energy of state~$|a\rangle$ by 
$\delta E_a({\cal E},{\bf k},\bm{\hat \epsilon}_{})$. This Stark shift can be written as the sum of an electric-multipole shift $\delta E_a({\cal E},{\bf k},\bm{\hat \epsilon}_{})$, a paramagnetic shift $\delta E_a^{({\rm P})}({\cal E},{\bf k},\bm{\hat \epsilon}_{})$ and a diamagnetic shift $\delta E_a^{({\rm D})}({\cal E},{\bf k},\bm{\hat \epsilon}_{})$:
\begin{align}
\label{eq:deltaEa}
&\delta E_a({\cal E},{\bf k},\bm{\hat \epsilon}_{}) = \nonumber \\ 
& \qquad 
\delta E_a^{({\rm E})}({\cal E},{\bf k},\bm{\hat \epsilon}_{}) +\delta E_a^{({\rm P})}({\cal E},{\bf k},\bm{\hat \epsilon}_{}) +
\delta E_a^{({\rm D})}({\cal E},{\bf k},\bm{\hat \epsilon}_{})
\end{align}
with \cite{Beloy2025}
\begin{subequations}
\label{eq:deltas}
\begin{align}
&\delta E_a^{({\rm E})}({\cal E},{\bf k},\bm{\hat \epsilon}_{}) = ({\cal E}/2)^2 \times \nonumber \\
&\quad \sum_p \left(
\frac{\langle a | v_{\rm E} | p \rangle \langle p | v_{\rm E}^\dagger | a \rangle}{\omega_{ap} - \omega} 
+ \frac{\langle a | v_{\rm E}^\dagger | p \rangle \langle p | v_{\rm E} | a \rangle}{\omega_{ap} + \omega}
\right),
\label{eq:Eshift}
\\
&\delta E_a^{({\rm P})}({\cal E},{\bf k},\bm{\hat \epsilon}_{}) = ({\cal E}/2c)^2 \times \nonumber \\
&\quad \sum_p \left(
\frac{\langle a | v_{\rm M} | p \rangle \langle p | v_{\rm M}^\dagger | a \rangle}{\omega_{ap} - \omega} 
+ \frac{\langle a | v_{\rm M}^\dagger | p \rangle \langle p | v_{\rm M} | a \rangle}{\omega_{ap} + \omega}
\right)
\label{eq:Pshift}
\end{align}
and
\begin{align}
&\delta E_a^{({\rm D})}({\cal E},{\bf k},\bm{\hat \epsilon}_{}) = ({\cal E}/2c)^2 \langle a | v_{\rm D} | a \rangle.
\label{eq:Dshift}
\end{align}
\end{subequations}
In these equations,
\begin{subequations}
\begin{align}
v_{\rm E} &\equiv \int_0^1 {\bf r}\cdot {{\bm {\hat \epsilon}}}_{} \exp(i\lambda {\bf k}\cdot {\bf r})\, {\rm d}\lambda, \label{eq:vE}\\
v_{\rm M} &\equiv \frac{i}{2} \int_0^1 \left(
\nabla \cdot \left[{\bf r}{\bm \times} {\bm {\hat \beta}}_{} \exp(i\lambda {\bf k}\cdot{\bf r})\right] \right. \nonumber \\
& \qquad \;\;+ \left. \left[{\bf r}{\bm \times} {\bm {\hat \beta}}_{} \exp(i\lambda {\bf k}\cdot{\bf r})\right] \cdot \nabla\right) \,\lambda\,{\rm d}\lambda,\\
v_{\rm D} &\equiv \left({\bf r}{\bm \times} {\bm {\hat \beta}}_{}\right)\cdot \left({\bf r}{\bm \times} {\bm {\hat \beta}}^*_{}\right) \times \nonumber\\ 
& \qquad \int_0^1 \int_0^1  \exp[i(\lambda-\lambda')\,{\bf k}\cdot{\bf r}\,] \, \lambda \,\lambda' \,{\rm d}\lambda\,{\rm d}\lambda',
\label{eq:vD}
\end{align}
\end{subequations}
and
\begin{equation}
    \omega_{ap} = E_a - E_p.
    \label{eq:omegaap}
\end{equation}
The interaction $v_{\rm M}$ can also be written in the following form,
\begin{align}
v_{\rm M} &\equiv \frac{1}{2} \int_0^1 \left[
{\bf L} \cdot{\bm {\hat \beta}}_{} \,\exp(i\lambda kz) 
 + \exp(i\lambda kz) \,{\bm {\hat \beta}}_{} \cdot{\bf L}\right] \,\lambda\,{\rm d}\lambda,
\end{align}
where ${\bf L}$ is the orbital angular momentum operator.

Let us assume that state $|a\rangle$ has a well defined magnetic quantum number $m_a^{\rm L}$ in a system of axes, the L~system, which is fixed with respect to the laboratory.
As shown by Eqs.~(\ref{eq:deltaEa}) to (\ref{eq:omegaap}),
\begin{equation}
\delta E_a({\cal E},{\bf k},\bm{\hat \epsilon}_{}) = \left(\frac{\cal E}{2}\right)^2\langle n_a l_a m_a^{\rm L} | {\cal V}({\bf k},\bm{\hat \epsilon}_{}) | n_a l_a m_a^{\rm L}\rangle,
\end{equation}
where ${\cal V}({\bf k},\bm{\hat \epsilon}_{})$ is a certain operator.
Let us also consider another system of axes, the S~system, whose $z$-axis is along the wave vector ${\bf k}$ and whose $xy$-plane is subtended by the vectors $\bm{\hat \epsilon}$ and $\bm{\hat \beta}$. The L~system can be brought to the S~system by a rotation with Euler angles $\alpha$, $\beta$ and $\gamma$. This rotation would transform $|n_a l_a m_a^{\rm L}\rangle$ into a linear combination of eigenvectors $|n_a l_a m_a\rangle$ of the $z$-component of ${\bf L}$ in the S-system. In terms of Wigner rotation matrices,
\begin{equation}
|n_a l_a m_a^{\rm L} \rangle = \sum_{m_a} {\cal D}_{m_a^{\rm L} m_a}^{\,(l_a)}(\alpha,\beta,\gamma)\, |n_a l_a m_a\rangle.
\end{equation}
Therefore
\begin{align}
&\delta E_a({\cal E},{\bf k},\bm{\hat \epsilon}_{}) = \nonumber \\ & \quad \left(\frac{\cal E}{2}\right)^2 \sum_{m_a' m_a} {\cal D}_{m_a^{\rm L} m_a'}^{\,(l_a)*}(\alpha,\beta,\gamma) {\cal D}_{m_a^{\rm L} m_a}^{\,(l_a)}(\alpha,\beta,\gamma) \times \nonumber \\
&\qquad \qquad \qquad \qquad\langle n_a l_a m_a' | {\cal V}({\bf k},\bm{\hat \epsilon}_{}) | n_a l_a m_a\rangle.
\end{align}

Assuming that the BBR field is both unpolarized and isotropic, the BBR shift of state $|a\rangle$, $\delta E_a$, can then be obtained by averaging $\delta E_a({\cal E},{\bf k},\bm{\hat \epsilon}_{})$ over the three Euler angles and over the frequency spectrum of the BBR field. To this end, we equate the power density $({\cal E}/2)^2/2\pi$ to $u(\omega,T)\,{\rm d}\omega$ \cite{Beloy2025}, with
\begin{equation}
    u(\omega,T) = \frac{1}{\pi^2 c^3} \frac{\omega^3}{\exp(\omega/k_{\rm B}T) -1},
\end{equation}
and write
\begin{align}
&\delta E_a = \frac{2}{\pi c^3} \int_0^\infty \frac{\omega^3\,{\rm d}\omega}{\exp(\omega/k_{\rm B}T) -1} \times \nonumber \\
&\qquad \frac{1}{8\pi^2} \int_0^{2\pi} {\rm d}\alpha \int_0^\pi {\rm d}\beta \sin\beta \int_0^{2\pi} {\rm d}\gamma \, \delta E_a({\cal E},{\bf k},\bm{\hat \epsilon}_{}).
\end{align}
Integrating over $\alpha$, $\beta$ and $\gamma$ collapses the sum over $m_a'$ into a single term due to the orthogonality of the rotation matrices. The result is
\begin{align}
&\delta E_a = \frac{2}{\pi c^3} \int_0^\infty \frac{\omega^3\,{\rm d}\omega}{\exp(\omega/k_{\rm B}T) -1} \times \nonumber \\ &\qquad \qquad \frac{1}{2l_a + 1}\sum_{m_a}
\langle n_a l_a m_a | {\cal V}({\bf k},\bm{\hat \epsilon}_{}) | n_a l_a m_a\rangle,
\label{eq:shiftgeneral}
\end{align}
where $\bm{\hat \epsilon}$ is an arbitrary unit vector in the $xy$-plane and it is understood that the vector ${\bf k}$ is in the positive $z$-direction.
As expected since the BBR field is isotropic, $\delta E_a$ does not depend on $m_a^{\rm L}$.

As above, the BBR shift $\delta E_a$ is the sum of an electric-multipole shift $\delta E_a^{({\rm E})}$, a paramagnetic shift $\delta E_a^{({\rm P})}$ and a diamagnetic shift $\delta E_a^{({\rm D})}$:
\begin{equation}
   \delta E_a = \delta E_a^{({\rm E})} + \delta E_a^{({\rm P})} + \delta E_a^{({\rm D})}.
\end{equation}
We consider each of these three terms in the next sections and in appendices \ref{section:appendixE}, \ref{section:appendixP} and \ref{section:appendixD}. The details of the calculation depend on the choice of polarization vector $\bm{\hat \epsilon}_{}$ but the final results do not.

\subsection{Electric-multipole shift}
\label{section:Eret}

The calculation of $\delta E_a^{({\rm E})}$ is outlined in Appendix~\ref{section:appendixE}. In view of the results of Sec.~\ref{section:mainsection}, it can be expected that for a Rydberg state $|a\rangle$, $\delta E_a^{({\rm E})}$ is dominated by the contribution of intermediate states $|p\rangle$ close in energy to this state and such that $|E_p - E_a|/(k_{\rm B}T) \ll 1$. Retaining only the term of leading order in $|E_p - E_a|/(k_{\rm B}T)$ in the calculation amounts to approximating $\delta E_a^{({\rm E})}$ by $\delta E_a^{({\rm E})\prime}$, where $\delta E_a^{({\rm E})\prime}$ is given by Eqs.~(\ref{eq:48tilde}) and (\ref{eq:deltaEEtilde}). The right-hand side of this last equation can be recast as a simpler series of powers in $k_{\rm B}T/c$, namely
\begin{equation}
\delta E_a^{({\rm E})\prime} = \sum_{n=1}^\infty c_{n}^{({\rm E})} \frac{(k_{\rm B}T)^{2n}}{c^{2n+1}}\,
\langle n_a l_a | r^{2n-2} | n_a l_a\rangle.
\label{eq:deltaEEexp}
\end{equation}
The coefficients of this series do not depend on $T$, $n_a$ or $l_a$ and have the following general form,
\begin{equation}
    c_{n}^{({\rm E})} = \frac{1}{\pi}\, \left(a_{n}^{({\rm E})}/b_{n}^{({\rm E})}\right)\,\zeta(2n),
    \label{eq:deltaEEexp2}
\end{equation}
where $a_{n}^{({\rm E})}$ and $b_{n}^{({\rm E})}$ are integers \cite{DRO}.
\renewcommand{\arraystretch}{1.2}

The BBR shift $\delta E_a^{({\rm E})\prime}$ thus depends on the temperature in a remarkably simple way.
The leading term in its expansion in powers of $(k_{\rm B}T)/c$ is the asymptotic BBR shift $\delta E_a^{({\rm E}1)\prime}$ found in the dipole approximation \cite{Farley1981,Gallagher1979} ($c_1^{({\rm E})} = \pi/3$). The next term is the sum of the partial shifts $\delta E_{a,00}^{({\rm E}2)\prime}$, $\delta E_{a,10}^{({\rm E}1)\prime}$ and $\delta E_{a,01}^{({\rm E}1)\prime}$ of Eq.~(\ref{eq:48tilde}), i.e., the electric-quadrupole shift $\delta E^{({\rm E}2)\prime}_a$ of Eq.~(\ref{eq:E2andD}) and the leading-order retardation corrections to the electric-dipole shift. These three terms contribute, respectively, $\pi^3/45$, $-\pi^3/135$ and $-\pi^3/135$ to the coefficient $c_2^{({\rm E})}$, which gives $c_2^{({\rm E})} = \pi^3/135$. The latter two terms thus largely cancel the electric-quadrupole shift $\delta E_a^{({\rm E}2)}$.

Calculating $\delta E_a^{({\rm E})\prime}$ by direct summation of the right-hand side of Eq.~(\ref{eq:deltaEEexp}) may be problematic because the matrix elements $\langle n_a l_a | r^{2n-2} | n_a l_a\rangle$ increase exponentially when $n_a \rightarrow \infty$ while the factors multiplying these matrix elements decrease only slowly. These two expansions are divergent and asymptotic. The difficulty
can be traced to Eq.~(\ref{eq:deltaEEexp0}), which formulates $\delta E_a^{({\rm E})\prime}$ as a sum of integrals of $u(\omega,T)$ multiplied by a power of $\omega$ rather than as the integral of $u(\omega,T)$ multiplied by a function of $\omega$. It can be avoided by using the fact that
\begin{equation}
 \int_0^\infty \frac{\omega^{2n-1}\,{\rm d}\omega}{\exp(\omega/k_{\rm B}T) -1} =  (2n-1)! \,\zeta(2n)\,(k_{\rm B}T)^{2n}
\end{equation}
and write Eq.~(\ref{eq:deltaEEexp}) in the following form, which does not involve an exponentially divergent series:
\begin{align}
\delta E_a^{({\rm E})\prime} &= \int_0^\infty 
\frac{{\rm d}\omega}{\exp(\omega/k_{\rm B}T)-1}\,\times \nonumber \\ 
&\quad \frac{1}{\pi}\,\sum_{n=1}^\infty \frac{a_{n}^{({\rm E})}/b_{n}^{({\rm E})}}{(2n-1)!}\, \frac{\omega^{2n-1}}{c^{2n+1}}
\,\langle n_a l_a | r^{2n-2} | n_a l_a\rangle.
\label{eq:deltaEEexp3}
\end{align}

\subsection{Paramagnetic shift}
\label{section:Pret}

The paramagnetic shift without retardation is normally negligible for Rydberg states \cite{Beloy2025}.
The paramagnetic shift with retardation, $\delta E_a^{({\rm P})}$, can be calculated in the same way as the electric shift $\delta E_a^{({\rm E})}$. Like the latter, we can approximate $\delta E_a^{({\rm P})}$ by a shift $\delta E_a^{({\rm P})\prime}$, assuming that the states and temperatures considered are such that $\delta E_a^{({\rm P})}$ is dominated by the contribution of intermediate states for which $|E_p - E_a| \ll k_B T$. Proceeding as above yields $\delta E_a^{({\rm P})\prime}$ as a power series of the following form,
\begin{equation}
\delta E_a^{({\rm P})\prime} = \sum_{n=2}^\infty 
C_{n}^{({\rm P})}(n_a,l_a) \frac{(k_{\rm B}T)^{2n}}{c^{2n+3}},
\end{equation}
where the coefficients $C_{n}^{({\rm P})}(n_a,l_a)$ vary from state to state but do not depend on $T$. Each term in this series is a factor $1/c^2$ smaller than the term of the same order in $k_{\rm B}T$ in the expansion of $\delta E_a^{({\rm E})\prime}$. The paramagnetic shift can therefore be expected to be very small compared to the electric-multipole shift even when retardation is taken into account. Moreover, as each of the terms in the expansion of $\delta E_a^{({\rm P})\prime}$ is of the same order in $1/c$ as the relativistic corrections neglected in the present calculations, these corrections would also need to be taken into account in a consistent treatment of the BBR shift to that order in $1/c$.

The coefficients $C_{n}^{({\rm P})}(n_a,l_a)$ are worked out in Appendix~\ref{section:appendixP} for the specific case of s-states ($l_a = 0$). The calculation results in a power series similar to that of Eq.~(\ref{eq:deltaEEexp}):
\begin{equation}
\delta E_a^{({\rm P})} = \sum_{n=2}^\infty c_{n}^{({\rm P})} \frac{(k_{\rm B}T)^{2n}}{c^{2n+3}}
\,\langle n_a l_a | r^{2n-2} | n_a l_a\rangle
\label{eq:deltaEPexp}
\end{equation}
with
\begin{equation}
    c_{n}^{({\rm P})} = \frac{1}{\pi}\, \left(a_{n}^{({\rm P})}/b_{n}^{({\rm P})}\right)\,\zeta(2n),
    \label{eq:deltaEPexp2}
\end{equation}
where $a_{n}^{({\rm P})}$ and $b_{n}^{({\rm P})}$ are integers 
\cite{DRO}.
However, the calculation does not straightforwardly generalizes to non-s states.

\subsection{Diamagnetic shift}
\label{section:Dret}

The calculation of $\delta E_a^{({\rm D})}$ is outlined in Appendix~\ref{section:appendixD}. This shift takes on a form similar as that found for the electric-multipole shift $\delta E_a^{({\rm D})\prime}$, i.e.,
\begin{equation}
\delta E_a^{({\rm D})} = \sum_{n=2}^\infty c_{n}^{({\rm D})} \frac{(k_{\rm B}T)^{2n}}{c^{2n+1}}
\,\langle n_a l_a | r^{2n-2} | n_a l_a\rangle.
\label{eq:deltaEDexp}
\end{equation}
The coefficients $c_{n}^{({\rm D})}$ are easily calculated.
Like the coefficients $c_{2n}^{({\rm E})}$ of the expansion of the shift $\delta E_a^{({\rm E})\prime}$, they do not depend on $T$, $n_a$ or $l_a$. They also have the same general form,
\begin{equation}
    c_{n}^{({\rm D})} = \frac{1}{\pi}\, \left(a_{n}^{({\rm D})}/b_{n}^{({\rm D})}\right)\,\zeta(2n),
    \label{eq:deltaEDexp2}
\end{equation}
where $a_{n}^{({\rm D})}$ and $b_{n}^{({\rm D})}$ are integers 
\cite{DRO}.
The first term in this series is of order $(k_{\rm B}T)^4$ and has $c_2^{({\rm D})} = \pi^3/45$, in complete agreement with Eq.~(\ref{eq:dia}). The higher order terms originate from retardation. 

The right-hand side of Eq.~(\ref{eq:deltaEDexp}) is an asymptotic series, like the right-hand side of Eq.~(\ref{eq:deltaEEexp}). Calculations of the diamagnetic shift requiring a summation of this series to a large number of terms may thus be best based on the following equation, which avoids the exponential divergence affecting Eq.~(\ref{eq:deltaEDexp}):
\begin{align}
\delta E_a^{({\rm D})} &= \int_0^\infty 
\frac{{\rm d}\omega}{\exp(\omega/k_{\rm B}T)-1}\,\times \nonumber \\ 
&\quad \frac{1}{\pi}\,\sum_{n=1}^\infty \frac{a_{n}^{({\rm D})}/b_{n}^{({\rm D})}}{(2n-1)!}\, \frac{\omega^{2n-1}}{c^{2n+1}}
\,\langle n_a l_a | r^{2n-2} | n_a l_a\rangle.
\label{eq:deltaEDexp3}
\end{align}
This shift is always positive, as can be seen from the structure of Eq.~(\ref{eq:deltaED}).

\subsection{Total BBR shift}
\label{section:total}

It follows from the above that the BBR shift of a Rydberg state $|a\rangle$ is closely approximated by the sum of the electric-multipole shift $\delta E_a^{({\rm E})\prime}$ and the diamagnetic shift $\delta E_a^{({\rm D})}$:
\begin{equation}
\delta {E}_a \approx \delta E_a^{({\rm E})\prime} + \delta E_a^{({\rm D})}.
\label{eq:deltaEtot1}
\end{equation}
Thus, from Eqs.~(\ref{eq:E1tilde}), (\ref{eq:deltaEEexp3}) and (\ref{eq:deltaEDexp3}),
\begin{equation}
\delta {E}_a \approx \delta {E}^{\rm d}_a + \delta E^{\rm nd}_a
\label{eq:deltaEtot2}
\end{equation}
where $\delta {E}^{\rm d}_a$ is the asymptotic electric-dipole shift \cite{Farley1981,Gallagher1979}, 
\begin{equation}
\delta {E}^{\rm d}_a = \delta E^{({\rm E}1)\prime}_a = \frac{\pi}{3}\, \frac{(k_{\rm B}T)^2}{c^3},
\label{eq:deltaEdip}
\end{equation}
and $\delta {E}^{\rm nd}_a$ is a non-dipole contribution to the BBR shift: as an integral,
\begin{align}
\delta &E_a^{{\rm nd}} = 
\int_0^\infty 
\frac{\omega\,{\rm d}\omega}{\exp(\omega/k_{\rm B}T)-1} 
\, \times \nonumber \\ &
\frac{1}{\pi}\,\sum_{n=2}^\infty \left(\frac{a_{n}^{({\rm E})}}{b_{n}^{({\rm E})}} + \, \frac{a_{n}^{({\rm D})}}{b_{n}^{({\rm D})}} \right)
\frac{\omega^{2n-2}}{c^{2n+1}}
\,\frac{\langle n_a l_a | r^{2n-2} | n_a l_a\rangle}{(2n-1)!},
\label{eq:deltaEnd1}
\end{align}
or as an asymptotic series,
\begin{align}
&\delta E^{\rm nd}_a = \nonumber \\
&\qquad \sum_{n=2}^\infty \left(c_{n}^{({\rm E})} + c_{n}^{({\rm D})}\right) \frac{(k_{\rm B}T)^{2n}}{c^{2n+1}}
\,\langle n_a l_a | r^{2n-2} | n_a l_a\rangle.
\label{eq:deltaEnd2}
\end{align}
Taking into account only the first term in this last series gives an approximation of order $(k_{\rm B}T)^4$ to the non-dipole shift, namely 
\begin{align}
\delta E^{(2)}_a &= \frac{8\pi^3}{135}\,\frac{(k_{\rm B}T)^4}{c^5}\,
\langle a | r^2 | a \rangle.
\label{eq:E2}
\end{align}
Keeping the first two terms gives an approximation of order $(k_{\rm B}T)^6$:  
\begin{align}
\delta E^{(3)}_a &=
\frac{8\pi^3}{135}\,\frac{(k_{\rm B}T)^4}{c^5}\,
\langle a | r^2 | a \rangle 
-\frac{44\pi^5}{42525}\,\frac{(k_{\rm B}T)^6}{c^7}\,
\langle a | r^4 | a \rangle.
\label{eq:E3}
\end{align}

\begin{figure}[t!]
\centering
\includegraphics[width=\columnwidth]{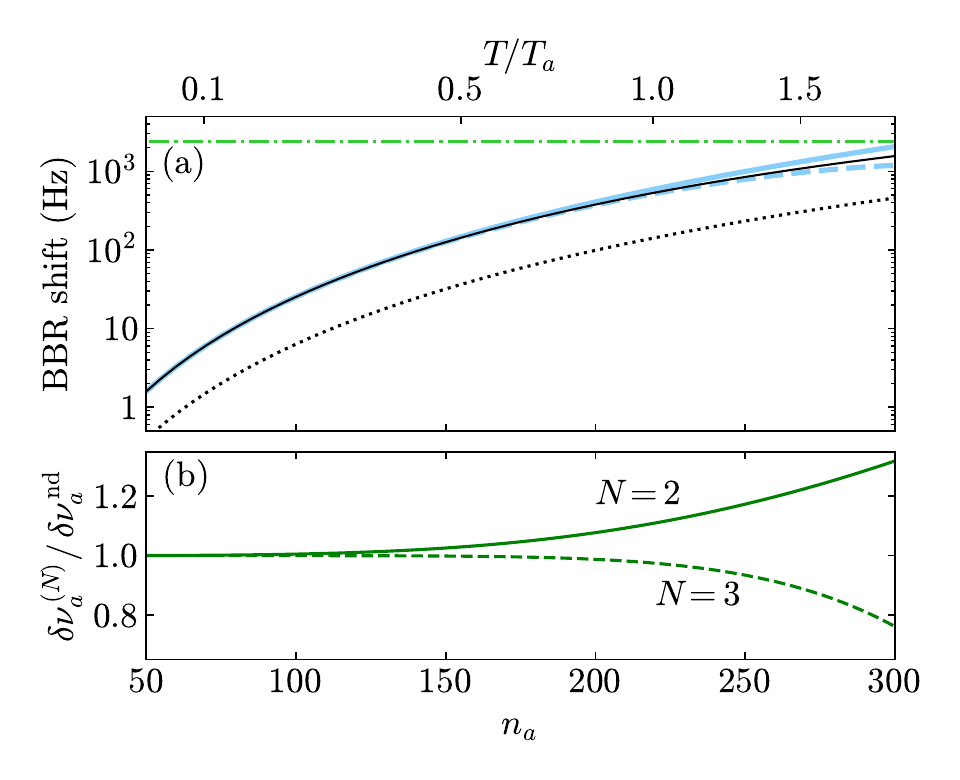}
\caption{
(a) The BBR shift of s-states of hydrogen at 300~K vs.\ the principal quantum number of the state. Solid black curve: the total non-dipole shift, ${\delta} \nu^{\rm nd}_a$. Dotted black curve: the contribution of the electric-multipole shift to the total non-dipole shift.  Solid blue curve: the non-dipole shifts $\delta \nu^{(2)}_a$. Dashed blue curve: the non-dipole shift $\delta \nu^{(3)}_a$. Dash-dotted line: the dipole shift, $\delta \nu^{{\rm d}}_a$. 
(b) The ratio of either the ${\delta} \nu^{(2)}_a$ shift (upper curve) or the ${\delta} \nu^{(3)}_a$ shift (lower curve) to the total non-dipole shift for the same states and temperature as in (a).
}
\label{fig:Fig2}
\end{figure}
\begin{figure}[t!]
\centering
\includegraphics[width=\columnwidth]{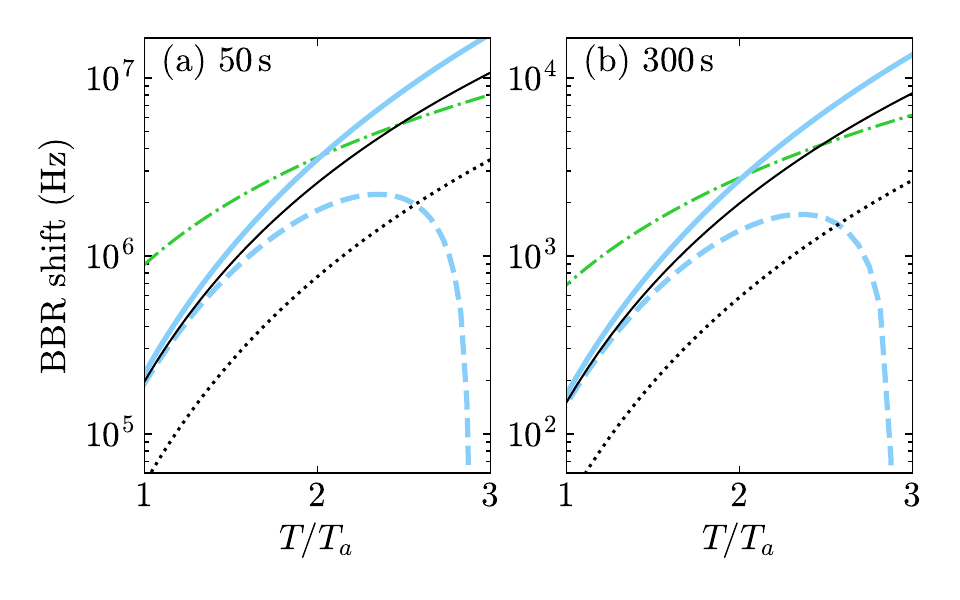}
\caption{
As in Fig,~\ref{fig:Fig2}(a), but here for the variation of the BBR shift with temperature for either the $50\,{\rm s}$ or $300\,{\rm s}$ state. The temperature is normalized to the characteristic temperature for the onset of retardation effect, $T_a$ ($T_a = 5770$~K for the $50\,{\rm s}$ state and 160~K for the $300\,{\rm s}$ state).
}
\label{fig:Fig3}
\end{figure}
\begin{figure}[t!]
\centering
\includegraphics[width=\columnwidth]{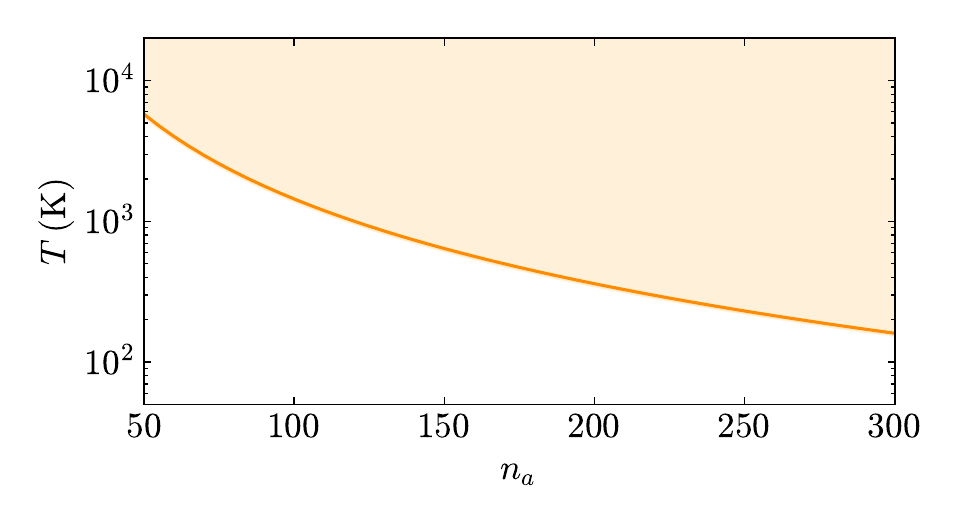}
\caption{
Solid curve: the characteristic temperature for the onset of retardation effects, $T_a$. Shaded area: the temperature region in which the total non-dipole shift ${\delta} E^{\rm nd}_a$ differs from ${\delta} E^{(2)}_a$ by more than 10\%.
}
\label{fig:Fig4}
\end{figure}

Numerical values of the non-dipole BBR shift $\delta E_a^{{\rm nd}}$ of s-states of hydrogen are presented in Figs.~\ref{fig:Fig2}, \ref{fig:Fig3} and \ref{fig:Fig4}.
We used the recursion relation \cite{Pasternack1937}
\begin{align}
&\langle n_al_a | r^k | n_al_a \rangle = \frac{(2k+1)\,n_a^2}{k+1} \langle n_al_a | r^{k-1}| n_al_a \rangle \nonumber\\ & \quad 
- \frac{k\,n_a^2}{4(k+1)}\left[(2l_a+1)^2-k^2\right]\langle n_al_a | r^{k-2}| n_al_a\rangle
\label{eq:Pasternack}
\end{align}
to calculate the required radial matrix elements, 
starting with
$\langle n_al_a | n_al_a \rangle = 1$ and
\begin{equation}
\langle n_al_a | r | n_al_a \rangle = [3n_a^2-l_a(l_a+1)]/2.
\end{equation}
We used both Eq.~(\ref{eq:deltaEnd1}) and (\ref{eq:deltaEnd2}) to calculate $\delta E_a^{{\rm nd}}$. 
In the case of Eq.~(\ref{eq:deltaEnd1}), we used Pad\'e approximants for summing up the series \cite{Wynn1966}. Obtaining all the results presented in the figures required including up to the first 44 terms of the series in the summation. Numerically integrating the result over $\omega$ presented no difficulty.

Figs.~\ref{fig:Fig2} and \ref{fig:Fig3} illustrate the growing importance of the higher-order terms in the expansion of the full non-dipole shift for increasing temperatures and increasing principal quantum numbers. The temperature $T_a$ appearing in these figures is the characteristic temperature for the onset of retardation defined by Eq.~(\ref{eq:Ta}).  Fig.~\ref{fig:Fig2} shows how the non-dipole BBR shift at $T = 300$~K varies with $n_a$, whereas Fig.~\ref{fig:Fig3} shows how the non-dipole BBR shift of the $50\,{\rm s}$ and $300\,{\rm s}$ states varies with $T$. As in Fig.~\ref{fig:Fig1}, we report the energy shifts of interest in terms of the corresponding frequency shifts --- thus in terms of the frequency shifts ${\delta} \nu^{\rm nd}_a$, $\delta \nu^{(2)}_a$, $\delta \nu^{(3)}_a$ and ${\delta} \nu^{\rm d}_a$ rather than in terms of the energy shifts
${\delta} E^{\rm nd}_a$, $\delta E^{(2)}_a$, $\delta E^{(3)}_a$ and ${\delta} E^{\rm d}_a$.
The non-dipole shifts ${\delta} \nu^{\rm nd}_a$, $\delta \nu^{(2)}_a$ and $\delta \nu^{(3)}_a$ are represented, respectively, by the solid black curves, solid blue curves and dashed blue curves. The dipole shift $\delta \nu_a^{{\rm d}}$ is represented by the dash-dotted green curves and the contribution of the electric-multipole shift to ${\delta} \nu^{\rm nd}_a$ by the dotted black curves. We note that the electric-multipole shift contributes only to between a quarter and a third of the full non-dipole shift, at least at the temperatures considered. The latter is thus dominated by the diamagnetic shift.

As shown by these results, the ratio of the temperature to the state-dependent characteristic temperature $T_a$ is a good indicator of the importance of the higher non-dipole corrections: $\delta \nu_a^{{\rm nd}} \approx \delta \nu^{(2)}_a$ when $T/T_a \ll 1$, whereas ${\delta} \nu^{\rm nd}_a$ departs significantly from $\delta \nu^{(2)}_a$ when $T/T_a \gg 1$. The temperature $T_a$ is practically identical to the temperature above which $\delta \nu_a^{{\rm nd}}$ differs from $\delta \nu_a^{(2)}$ by more than 10\% (Fig.~\ref{fig:Fig4}). 

Including the next-to-leading term in the calculation, as in Eq.~(\ref{eq:E3}), gives a value of the non-dipole shift differing from the full non-dipole shift by more than 10\% only above $T/T_a \approx 1.5$ --- e.g., compare the dashed blue curves to the black solid curves in Figs.~\ref{fig:Fig2}(a) and \ref{fig:Fig3}, and see Fig.~\ref{fig:Fig2}(b). The difference increases rapidly with $T$, particularly above $2\, T_a$. The series (\ref{eq:deltaEnd2}) strongly diverges at these high temperatures. The full non-dipole shift can still be calculated from Eq.~(\ref{eq:deltaEnd1}), though. As shown by Fig.~\ref{fig:Fig3}, it remains positive and continues to increase monotonically above $2\, T_a$, at least up to $T = 3\, T_a$. The non-dipole shift becomes larger than the dipole shift $\delta \nu_a^{{\rm d}}$ at $T \approx 2.5\, T_a$. The BBR shift is thus dominated by non-dipole couplings at still higher temperatures.

\section{Conclusions}
\label{secrtion:conclusions}

In summary, we have shown that the electric field of the thermal radiation contributes a shift identical in structure and of the same order in $k_{\rm B}T/c$ as the diamagnetic BBR shift of Eq.~(\ref{eq:dia}). Albeit smaller in magnitude, this contribution may need to be taken into account in circumstances where the diamagnetic shift is relevant. More generally, we have studied how the familiar electric-dipole formulation of the BBR shift can be extended to encompass retardation so as to make it possible to explore higher temperature regimes. The calculations are largely based on the observation that the electric-multipole shifts are dominated by the contribution of intermediate states close in energy to the state of interest. This makes it possible to replace the functions $F_K(y)$ and $F_{Kqq'}(y)$ of Eqs.~(\ref{eq:FK}) and (\ref{eq:FKqqprime}) by their small-$|y|$ forms. In turn, this makes it possible to reduce the electric-multipole shifts with retardation to a remarkably simple form. The diamagnetic shift with retardation can be similarly simplified. The approximation improves when the temperature increases, which is also when corrections to the electric-dipole BBR shift become relevant. Altogether, and as anticipated in Ref.~\cite{Farley1981}, the BBR shift needs to be calculated beyond the electric-dipole approximation at temperatures close or exceeding the characteristic temperature of Eq.~(\ref{eq:Ta}). For atomic hydrogen, the total BBR shift is dominated by non-dipole contributions at temperatures above $2.5~T_a$.

It would be of interest to extend these numerical calculations to other species, in particular to alkalis. We have shown that the sum rules used in the present work are exact in the case of single active electron systems described by an $l$- and $j$-dependent model potential. The results for cesium presented in Sec.~\ref{section:Cesium} indicate that these sum rules and  the dominance of neighboring intermediate states also hold for for the electric-quadrupole shift of alkali metals. The above results for the full BBR shift of hydrogen are thus likely to extend to these more complex systems, at least qualitatively. However, a calculation of the higher-order BBR shifts would be necessary for reaching a definitive conclusion on this issue.

\begin{center}
{\small \bf DATA AVAILABILITY}
\end{center}

The data that support the findings of this article are openly available \cite{DRO}, embargo periods may apply.

\appendix
\section{Sum rules used in this work} 
\label{section:appendix}

The following sum rules underpin many of the calculations outlined in the body of the text and in appendices \ref{section:appendixE} and \ref{section:appendixP}:
\begin{align}
&\sum_{n_p} (E_p - E_a)
\langle a | r^{k_1} | p\rangle\langle p | r^{k_2} | a\rangle = \nonumber \\ & \quad  \frac{1}{2}\,[k_1k_2 + l_p(l_p + 1) - l_a(l_a+1)] \langle a | r^{k_1 + k_2 -2} | a \rangle
\label{eq:sum0}
\end{align}
and
\begin{align}
&\sum_{n_pl_p} (2l_p + 1) \begin{pmatrix} l_a & K & l_p \\ 0 & 0 & 0 \end{pmatrix}^2 (E_p - E_a)
\langle a | r^{k_1} | p\rangle\langle p | r^{k_2} | a\rangle \nonumber \\ & \qquad \qquad  = \frac{1}{2}\,[k_1k_2 + K(K + 1)] \langle a | r^{k_1 + k_2 -2} | a \rangle.
\label{eq:sum}
\end{align}
More generally,
\begin{align}
&\sum_{n_p} (E_p - E_a) \langle a | f(r) | p\rangle\langle p | g(r) | a\rangle = \nonumber \\ &
\frac{1}{2}\,
\left\langle
a \left| \frac{{\rm d}f}{{\rm d}r}\frac{{\rm d}g}{{\rm d}r}
+ [l_p(l_p+1)-l_a(l_a+1)]\,
\frac{f(r)g(r)}{r^2} \right| a \right\rangle
\label{eq:sum_npK}
\end{align}
and
\begin{align}
\sum_{n_pl_p} & (2l_p + 1) \begin{pmatrix} l_a & K & l_p \\ 0 & 0 & 0 \end{pmatrix}^2 \nonumber\\ & \qquad \qquad \qquad  \times (E_p - E_a)
\langle a | f(r) | p\rangle\langle p | g(r) | a\rangle \nonumber \\ & \quad  = \frac{1}{2}\,
\left\langle
a \left| \frac{{\rm d}f}{{\rm d}r}\frac{{\rm d}g}{{\rm d}r}
+ K(K + 1)\,
\frac{f(r)g(r)}{r^2} \right| a \right\rangle,
\label{eq:sumfg}
\end{align}
where $f(r)$ and $g(r)$ are any differentiable functions of $r$.
Our proof of these results is based on the fact that the functions $R_a(r)$ and $R_p(r)$ are eigenfunctions of the corresponding radial Hamiltonians, $h_a$ and $h_p$ \cite{Jackiw1967}:
\begin{equation}
    h_a R_a(r) = E_a R_a(r), \qquad h_p R_p(r) = E_p R_p(r),
\end{equation}
with
\begin{subequations}
\begin{align}
h_a &\equiv -\frac{1}{2}\left[\frac{1}{r^2}\frac{{\rm d}\;}{{\rm d}r}r^2\frac{{\rm d}\;}{{\rm d}r} - \frac{l_a(l_a+1)}{r^2}\right] + V(r), \\
h_p & \equiv -\frac{1}{2}\left[\frac{1}{r^2}\frac{{\rm d}\;}{{\rm d}r}r^2\frac{{\rm d}\;}{{\rm d}r} - \frac{l_p(l_p+1)}{r^2}\right] + V(r), 
\end{align}
\end{subequations}
where $V(r)$ is a 1-electron potential.
Thus 
\begin{align}
(E_p - E_a) \langle p | g(r) | a\rangle &= \int_0^\infty \left[h_p R_p^*(r)\right] g(r) R_a(r) r^2 {\rm d}r 
\nonumber \\ 
&- \int_0^\infty
R_p^*(r) g(r) \left[h_a R_a(r)\right] r^2 \,{\rm d}r.
\end{align}
In view of the Hermiticity of the Hamiltonian $h_p$,
\begin{align}
(E_p - E_a) \langle p | g(r) | a\rangle &= \int_0^\infty R_p^*(r) \left[h_p \, g(r) R_a(r)\right] r^2 {\rm d}r 
\nonumber \\ 
&- \int_0^\infty
R_p^*(r) g(r) \left[h_a R_a(r)\right] r^2 \,{\rm d}r.
\end{align}
Using the completeness of the radial functions $R_p(r)$ then yields Eqs.~(\ref{eq:sum0}) and (\ref{eq:sum_npK}) after a short calculation.
We carry out the summation over $l_p$ by making use of the sum rules \cite{Pain2021}
\begin{align}
   & \sum_{j,m} (2j+1) j(j+1) \begin{pmatrix}    j_1 & j_2 & j \\ m_1 & m_2 & m \end{pmatrix}^2 = \nonumber \\ & \qquad \qquad \qquad   j_1(j_1+1) + j_2(j_2+1) + 2 m_1m_2
\end{align}
and \cite{Edmonds}
\begin{equation}
    \sum_{j,m} (2j+1) \begin{pmatrix}    j_1 & j_2 & j \\ m_1 & m_2 & m \end{pmatrix}^2 = 1,
\end{equation}
here with $m_1 = m_2 = 0$. Eqs.~(\ref{eq:sum}) and (\ref{eq:sumfg}) follow.

\section{Calculation of $\delta E_a^{({\rm E})}$}
\label{section:appendixE}

For simplicity, we assume here that the wave vector ${\bf k}$ is in the positive $z$-direction and that
\begin{equation}
    \bm{\hat \epsilon}_{} = \bm{\hat \epsilon}_{1} = -(\mathbf{\hat x} + i \mathbf{\hat y})/\sqrt{2},
\label{eq:epsilon}
\end{equation}
with $\mathbf{\hat x}$ and $\mathbf{\hat y}$ being unit vectors in the $x$- and $y$-directions.
Then, from Eqs.~(\ref{eq:Eshift}), (\ref{eq:vE}) and (\ref{eq:shiftgeneral}), 
\begin{align}
&\delta E_a^{({\rm E})} = \frac{2}{\pi c^3} \int_0^\infty \frac{\omega^3\,{\rm d}\omega}{\exp(\omega/k_{\rm B}T) -1} \times \nonumber \\
& \qquad \sum_{n_p l_p} \left[\frac{{\cal P}_-(n_a l_a n_p l_p,k)}{\omega_{ap}-\omega} + \frac{{\cal P}_+(n_a l_a n_p l_p,k)}{\omega_{ap}+\omega} \right]
\end{align}
with
\begin{subequations}
\begin{align}
&{\cal P}_-(n_a l_a n_p l_p,k) = \int_0^1 {\rm d}\lambda \int_0^1 {\rm d}\lambda' \,\frac{1}{2l_a+1} \times \nonumber \\
& \qquad \quad \sum_{m_a m_p} \langle n_a l_a m_a | {\bf r}\cdot\bm{\hat \epsilon}_{1} \exp(i\lambda kz) | n_p l_p m_p \rangle \times \nonumber \\ 
& \qquad \qquad\quad \;\;\langle n_p l_p m_p | {\bf r}\cdot\bm{\hat \epsilon}_{1}^* \exp(-i\lambda' kz) | n_al_a m_a\rangle
\label{eq:Pminus}
\end{align}
and
\begin{align}
&{\cal P}_+(n_a l_a n_p l_p,k) = \int_0^1 {\rm d}\lambda \int_0^1 {\rm d}\lambda' \,\frac{1}{2l_a+1} \times \nonumber \\
& \qquad \quad \sum_{m_a m_p} \langle n_a l_a m_a | {\bf r}\cdot\bm{\hat \epsilon}_{1}^* \exp(-i\lambda kz) | n_p l_p m_p \rangle \times \nonumber \\ 
& \qquad \qquad\quad \;\;\langle n_p l_p m_p | {\bf r}\cdot\bm{\hat \epsilon}_{1} \exp(i\lambda' kz) | n_al_a m_a\rangle.
\end{align}
\end{subequations}
We note that
\begin{equation}
\exp(i\lambda kz) = \sqrt{4\pi}\,\sum_{\Lambda=0}^\infty i^{\Lambda} \sqrt{2\Lambda + 1}\,j_{\Lambda}(\lambda k r)\,Y_{\Lambda 0}(\mathbf{\hat r})
\end{equation}
where the $j_{\Lambda}(\cdot)$'s are spherical Bessel functions. We also note 
that ${\bf r}\cdot\bm{\hat \epsilon}_{1} = (4\pi/3)^{1/2}\, r \,Y_{11}(\mathbf{\hat r})$ for our choice of polarization vector, and that
\begin{align}
&Y_{11}(\mathbf{\hat r}) Y_{\Lambda 0}(\mathbf{\hat r}) = \sum_{K \mu} \sqrt{\frac{3 (2\Lambda +1)(2K+1)}{4\pi}}\, \times \nonumber \\
&\qquad \qquad \qquad  \quad
\begin{pmatrix}
    1 & \Lambda & K \\ 1 & 0 & \mu 
\end{pmatrix} \begin{pmatrix}
    1 & \Lambda & K \\ 0 & 0 & 0 
\end{pmatrix} Y^*_{K\mu}(\mathbf{\hat r}).
\end{align}
Combining these results and using the recursion relation satisfied by the spherical Bessel functions leads to the following partial wave expansion for the product $Y_{11}(\mathbf{\hat r})\exp(i\lambda kz)$, 
\begin{align}
&Y_{11}(\mathbf{\hat r})\exp(i\lambda kz) = - \sqrt{\frac{3}{2}}\, \sum_{K=1}^\infty i^{K+1} \, \times \nonumber \\ & \qquad \qquad 
\sqrt{K(K+1)(2K+1)}\,\frac{j_{K}(\lambda kr)}{\lambda kr}\,Y_{K1}(\mathbf{\hat r}),
\end{align}
and similarly for the product  $Y^*_{11}(\mathbf{\hat r})\exp(-i\lambda' kz)$.
Therefore
\begin{widetext}
\begin{align}
&{\cal P}_-(n_a l_a n_p l_p,k) = -\frac{2\pi}{k^2} \int_0^1 {\rm d}\lambda \int_0^1 {\rm d}\lambda' \,\sum_{K K'} i^{K-K'} 
\frac{\sqrt{K K' (K+1) (K'+1) (2K+1) (2K'+1)}}{2l_a+1} \, \times \nonumber \\
& \qquad \qquad\frac{1}{\lambda \lambda'}\, \langle n_a l_a | j_K(\lambda kr) | n_p l_p \rangle 
\langle n_p l_p  | j_{K'}(\lambda' kr) | n_al_a \rangle 
\sum_{m_a m_p} \langle Y_{l_a m_a}|\, Y_{K1} |\, Y_{l_pm_p}\rangle \langle Y_{l_pm_p} |\, Y_{K'-1}|\, Y_{l_a m_a}\rangle.
\end{align}
The angular part of the right-hand side reduces to a single term since  \cite{Edmonds}
\begin{equation}
\sum_{m_a m_p} \begin{pmatrix}
    l_a & K & l_p \\ -m_a  & 1 & m_p 
\end{pmatrix} \begin{pmatrix}
    l_p & K' & l_p \\ -m_p  & -1 & m_a 
\end{pmatrix} \begin{pmatrix}
    l_a & K & l_p \\ 0  & 0 & 0 
\end{pmatrix} \begin{pmatrix}
    l_p & K' & l_p \\ 0 & 0 & 0
\end{pmatrix}= \frac{1}{2K+1} \begin{pmatrix}
    l_a & K & l_p \\ 0  & 0 & 0 
\end{pmatrix}^2
\,\delta_{K K'}.
\end{equation}
${\cal P}_+(n_a l_a n_p l_p,k)$ can be calculated in the same way, with the same result:
\begin{align}
&{\cal P}_{+}(n_a l_a n_p l_p,k) = {\cal P}_{-}(n_a l_a n_p l_p,k) = \frac{1}{2k^2}\int_0^1 {\rm d}\lambda \int_0^1 {\rm d}\lambda' 
\sum_{K = 1}^\infty (2K+1)K(K+1) \, \times \nonumber \\ 
& \qquad \qquad \qquad \qquad \qquad \qquad 
(2l_p+1)\begin{pmatrix}
    l_a & K & l_p \\ 0 & 0 & 0 
\end{pmatrix}^2
\frac{1}{\lambda \lambda'} \langle n_a l_a | j_K(\lambda kr) | n_p l_p\rangle \langle n_p l_p | j_K(\lambda' kr) | n_a l_a\rangle,
\label{eq:Ppm}
\end{align}
where $\langle n_a l_a | j_K (\lambda k r) | n_pl_p\rangle$ and $\langle n_p l_p | j_K (\lambda' k r) | n_al_a\rangle$ are the following radial matrix elements:
\begin{subequations}
\begin{align}
\langle n_a l_a | j_K (\lambda k r) | n_pl_p\rangle &= \int_0^\infty R_{n_al_a}^*(r) j_K(\lambda k r) R_{n_pl_p}(r)\,r^2\,{\rm d}r
,\\
\langle n_p l_p | j_K (\lambda' k r) | n_al_a\rangle &= \int_0^\infty R_{n_pl_p}^*(r) j_K(\lambda' k r) R_{n_al_a}(r)\,r^2\,{\rm d}r.
\end{align}
\end{subequations}
Expanding the spherical Bessel functions in powers of their argument yields 
\begin{equation}
\delta E_a^{({\rm E})} = \sum_{K=1}^\infty \sum_{q,q' = 0}^\infty  \delta E_{a,qq'}^{({\rm E}K)}
\label{eq:deltaEEexp0}
\end{equation}
with
\begin{align}
&\delta E_{a,qq'}^{({\rm E}K)} = - \frac{(-1)^{q+q'}}{q! q'!} \left(\frac{k_{\rm B}T}{c}\right)^{2K+2q+2q'+1} 
\sum_{n_p l_p} (2l_p+1) \begin{pmatrix}
    l_a & K & l_p \\ 0 & 0 & 0 
\end{pmatrix}^2 
F_{Kqq'}\left(\frac{E_p - E_a}{k_{\rm B}T}\right)
\times \nonumber \\ & \qquad \qquad \quad 
\qquad \qquad \qquad \qquad \qquad \qquad \qquad \qquad \qquad \qquad \qquad
\langle n_a l_a | r^{K+2q}|n_pl_p\rangle\langle n_p l_p | r^{K+2q'}| n_al_a\rangle.
\label{eq:48}
\end{align}
The function $F_{Kqq'}(y)$ is a generalization of the function $F_K(y)$ used in Sec.~\ref{section:mainsection}: $F_K(y) \equiv F_{K00}(y)$ and
\begin{align}
\label{eq:FKqqprime}
F_{Kqq'}(y) &= \frac{1}{\pi}\,\frac{K(K+1)(2K+1)}{2^{q+q'}(K+2q)(2K+2q+1)!!(K+2q')(2K+2q'+1)!!} \times \nonumber \\ & \qquad \qquad \qquad \qquad \qquad \qquad \qquad \qquad \qquad \qquad \qquad 
{\rm P.V.}\int_0^\infty\left(\frac{1}{y+x} + \frac{1}{y-x}\right)\frac{x^{2K+2q+2q'+1}}{\exp(x)-1}\,{\rm d}x\\
\label{eq:FKqqprimesmally}
&\sim -\frac{2y}{\pi}\,\frac{K(K+1)(2K+1)(2K+2q+2q'-1)!}{2^{q+q'}(K+2q)(2K+2q+1)!!(K+2q')(2K+2q'+1)!!}\,\zeta(2K+2q+2q') \quad \mbox{when $|y|\ll 1$.}
\end{align}
Replacing $F_{Kqq'}(y)$ by its small-$|y|$ form in Eq.~(\ref{eq:48}) and using Eq.~(\ref{eq:sum}) then leads to the following approximate partial shifts,
\begin{align}
\delta E_{a,qq'}^{({\rm E}K)\prime} &= \frac{1}{\pi}\,\frac{(-1)^{q+q'}}{q! q'!}  
\,\frac{K(K+1)(2K+1)(2K+2q+2q'-1)![(K+2q)(K+2q')+K^2+K]}{2^{q+q'}(K+2q)(2K+2q+1)!!(K+2q')(2K+2q'+1)!!} \,\times \nonumber \\
& \qquad \qquad \qquad \qquad \qquad \qquad \qquad \qquad \zeta(2K+2q+2q')\,\frac{(k_{\rm B}T)^{2K+2q+2q'}}{c^{2K+2q+2q'+1}}\,\langle n_a l_a | r^{2K+2q+2q'-2}| n_al_a\rangle,
\label{eq:48tilde}
\end{align}
and to the following approximate expression of the full electric-multipole shift,
\begin{equation}
\delta E_a^{({\rm E})} \approx  \delta E_a^{({\rm E})\prime} = \sum_{K=1}^\infty \sum_{q,q' = 0}^\infty \delta E_{a,qq'}^{({\rm E}K)\prime}.
\label{eq:deltaEEtilde}
\end{equation}
Comparing Eq.~(\ref{eq:deltaEKdefinednr}) to Eq.~(\ref{eq:48}) and Eq.~(\ref{eq:EKfinal2}) to Eq.~(\ref{eq:48tilde}) shows that $\delta E_{a,00}^{({\rm E}K)} \equiv \delta E_a^{({\rm E}K)}$ and $\delta E_{a,00}^{({\rm E}K)\prime} \equiv \delta E_a^{({\rm E}K)\prime}$. The present results thus reproduce the electric-multipole shifts of Sec.~\ref{section:mainsection} when retardation is neglected, as should be the case.

\section{Calculation of $\delta E_a^{({\rm P})}$ for s-states}
\label{section:appendixP}

Rather than taking the $z$-axis of the system of coordinates along ${\bf k}$, it is more convenient here to take the $x$-axis along ${\bf k}$ and set $\bm{\hat \beta}_{} = \mathbf{\hat z}$, where $\mathbf{\hat z}$ is a unit vector in the $z$-direction. Then,
for s-states ($l_a = 0$),
\begin{align}
&\langle a | {v}_{\rm M} | p \rangle \langle p | {v}_{\rm M}^\dagger | a \rangle = \frac{1}{4} \int_0^1 {\rm d}\lambda \lambda \int_0^1 {\rm d}\lambda' \lambda'
\, m_p^2\, \langle a| \exp(i\lambda kx) |p\rangle\langle p | \exp(-i\lambda' kx) | a\rangle.
\label{eq:prod1}
\end{align}
Using the expansion
\begin{equation}
\exp(i\lambda kx) = 4\pi\,\sum_{\Lambda=0}^\infty \sum_{\mu = -\Lambda}^\Lambda i^{\Lambda}j_{\Lambda}(\lambda k r)\,Y_{\Lambda\mu}^*(\mathbf{\hat x})Y_{\Lambda \mu}(\mathbf{\hat r})
\end{equation}
and integrating over the angles reduces Eq.~(\ref{eq:prod1}) to the following form,
\begin{align}
&\langle a | {v}_{\rm M} | p \rangle \langle p | {v}_{\rm M}^\dagger | a \rangle
= 4\pi^2 \int_0^1 {\rm d}\lambda \lambda \int_0^1 {\rm d}\lambda' \lambda'\,
 m_p^2 \left| Y_{l_pm_p}\left(\mathbf{\hat x}\right) \right|^2
\langle n_al_a| j_{l_p}(\lambda kr) |n_p l_p \rangle\langle n_p l_p | j_{l_p}(\lambda'kr) | l_a n_a\rangle,
\end{align}
and similarly for $\langle a | {v}_{\rm M}^\dagger | p \rangle \langle p | {v}_{\rm M} | a \rangle$.
The summation over intermediate states appearing in Eq.~(\ref{eq:Pshift}) involves a summation over $n_p$, a summation over $l_p$ and a summation over $m_p$. The latter reduces to summing $m_p^2|Y_{l_pm_p}(\mathbf{\hat x})|^2$ over the possible values of $m_p$. From general results \cite{Varshalovich},
\begin{equation}
\sum_{m_p} m_p^2 \left| Y_{l_pm_p}\left(\mathbf{\hat x}\right) \right|^2 = \frac{l_p (l_p+1) (2 l_p + 1)}{8\pi}.
\end{equation}
Expanding the spherical Bessel functions in power series, 
folding $\delta E_a^{({\rm P})}({\cal E},{\bf k},\bm{\hat \epsilon}_{})$ with the spectral and angular distributions of the BBR radiation, making the small $|E_p - E_a|/(k_{\rm B}T)$ approximation and summing over $n_p$ by using Eq.~(\ref{eq:sum0}) gives
\begin{align}
\delta E_a^{({\rm P})} \approx
\delta E_a^{({\rm P})\prime} &= \sum_{l_p = 0}^\infty \sum_{q,q'=0}^\infty
\frac{1}{4\pi}\,\frac{(-1)^{q+q'}}{q! q'!}\,
\frac{l_p(l_p+1)(2l_p+1) [(2q+l_p)(2q'+l_p)+l_p(l_p+1)] (2q+2q'+2l_p+1)!}
{(2l_p+2q+1)!(2l_p+2q'+1)!(2q+l_p+2)(2q'+l_p+2)} \nonumber \\
&\qquad \qquad \qquad \qquad \qquad \times
\zeta(2l_p+2q+2q'+2)\,\frac{(k_{\rm B}T)^{2l_p+2q+2q'+2}}{c^{2l_p+2q+2q'+5}}\, \langle n_al_a | r^{2l_p+2q+2q'+2} | n_a l_a\rangle.
\end{align}
We stress that this result only applies to the $l_a = 0$ case.

\section{Calculation of $\delta E_a^{({\rm D})}$}
\label{section:appendixD}

We take ${\bf k}$ take to be in the positive $z$-direction and set
\begin{equation}
    \bm{\hat \beta}_{} = -i\bm{\hat \epsilon}_{1} = (i\mathbf{\hat x} - \mathbf{\hat y})/\sqrt{2}.
\label{eq:beta}
\end{equation}
Thus
${\bf k}\cdot{\bf r} = kz = kr\cos\theta$, where $\theta$ is the angle between the vector ${\bf r}$ 
 and the positive $z$-axis, and 
\begin{align}
 \left({\bf r}{\bm \times} {\bm {\hat \beta}}_{}\right)\cdot \left({\bf r}{\bm \times} {\bm {\hat \beta}}^*_{}\right) &= \frac{1}{2}\,r^2 \,(1+\cos^2 \theta).
 \end{align}
Hence,
from Eqs.~(\ref{eq:Dshift}), (\ref{eq:vD}) and (\ref{eq:shiftgeneral}),
\begin{align}
&\delta E_a^{({\rm D})} = \frac{1}{\pi c^5} \int_0^\infty \frac{\omega^3\,{\rm d}\omega}{\exp(\omega/k_{\rm B}T) -1} \int_0^1 {\rm d}\lambda \int_0^1 {\rm d}\lambda' \,\frac{\lambda\,\lambda'}{2l_a + 1} 
\sum_{m_a} \left\langle a \left| 
\, r^2 \, (1 + \cos^2 \theta) 
\exp[i(\lambda-\lambda')\,kr\cos\theta\,]  \right| a \right\rangle.
\label{eq:first}
\end{align}

\noindent Since
 \begin{align}
 \cos^2 \theta\, \exp[{i(\lambda-\lambda')kr\cos\theta}] =
 \frac{{\rm d}\;\;\;\;\;}{{\rm d}{\lambda k r}}\,\exp(i \lambda k r \cos\theta)\,
 \frac{{\rm d}\;\;\;\;\;\,}{{\rm d}{\lambda'k r}}\exp(-i\lambda' k r \cos\theta),
 \end{align}
expanding the exponentials in partial waves yields
 \begin{align}
&\sum_{m_a}\left\langle a \left|  \, r^2\, (1 + \cos^2\theta)\, 
\exp[i(\lambda-\lambda')\,kr\cos\theta\,]  \right| a \right\rangle =
4\pi \sum_{\Lambda, \Lambda' = 0}^\infty i^{\Lambda - \Lambda'} \sqrt{(2\Lambda+1)(2\Lambda' + 1)} \, \times \nonumber \\ & \qquad \qquad \qquad \qquad \qquad 
\langle n_a l_a | \,r^2 j_{\Lambda}(\lambda kr) j_{\Lambda'}(\lambda' kr) + r^2 j_{\Lambda}^\prime(\lambda kr) j_{\Lambda'}^\prime(\lambda' kr)\,| n_a l_a\rangle \,\sum_{m_a}\langle Y_{l_a m_a}| \,Y_{\Lambda 0}Y_{\Lambda' 0}\,|\, Y_{l_am_a}\rangle,
\label{eq:second}
 \end{align}
where $j_n^\prime(\cdot)$ denotes 
the derivative of the spherical Bessel function $j_n(\cdot)$ with respect to its argument. 
We calculate the angular matrix element by writing $Y^*_{l_am_a}(\mathbf{\hat r})Y_{\Lambda 0}(\mathbf{\hat r})$ and $Y_{\Lambda' 0}(\mathbf{\hat r})Y_{l_am_a}(\mathbf{\hat r})$ as linear combinations of spherical harmonics, with the result that 
\begin{align}
\sum_{m_a}\langle Y_{l_a m_a}| \,Y_{\Lambda 0}Y_{\Lambda' 0}\,|\, Y_{l_am_a}\rangle &= 
\sum_{m_a}
\sum_{K \mu} \sum_{K'\mu'} (-1)^{m_a+\mu'}(2l_a+1)\frac{\sqrt{(2\Lambda +1)(2K+1)(2\Lambda'+1){2K'+1)}}}{4\pi}\, \times \nonumber \\
&\qquad \qquad
\begin{pmatrix}
    l_a & \Lambda & K \\ -m_a & 0 & \mu 
\end{pmatrix} \begin{pmatrix}
    l_a & \Lambda & K \\ 0 & 0 & 0 
\end{pmatrix}
\begin{pmatrix}
    \Lambda' & l_a & K' \\ 0 & m_a & \mu' 
\end{pmatrix} \begin{pmatrix}
    \Lambda' & l_a & K' \\ 0 & 0 & 0 
\end{pmatrix} \, \delta_{KK'} \delta_{\mu\, -\mu'}.
\label{eq:third}
\end{align}
Combining Eqs.~(\ref{eq:first}),  (\ref{eq:second}) and (\ref{eq:third}) then gives
\begin{align}
\delta E_a^{({\rm D})} = \frac{1}{\pi c^5} \int_0^\infty \frac{\omega^3\,{\rm d}\omega}{\exp(\omega/k_{\rm B}T) -1} &\int_0^1 {\rm d}\lambda \int_0^1 {\rm d}\lambda' \,{\lambda\,\lambda'}\,\times \nonumber \\ &  \qquad
\sum_{\Lambda = 0}^\infty (2\Lambda + 1)
\langle n_a l_a |\, r^2 j_{\Lambda}(\lambda kr) j_{\Lambda}(\lambda' kr) + r^2 j_{\Lambda}^\prime(\lambda kr) j_{\Lambda}^\prime(\lambda' kr)\,| n_a l_a\rangle.
\label{eq:deltaED} 
\end{align}
Expanding the Bessel functions in power series, replacing $k$ by $\omega/c$ 
and integrating the result over $\omega$, $\lambda$ and $\lambda'$  reduces the diamagnetic shift
to the form of Eq.~(\ref{eq:deltaEDexp}).
\end{widetext}



\begin{thebibliography}{99}
\bibitem{77K}
These multipole shifts decrease rapidly with $T$, though. For example, the quadrupole frequency shift of the $50\,{\rm s}$ state is 1.2~Hz at 300~K but only 5.2~mHz at 77~K. The temperature at which the quadrupole shift is of the order of 1~Hz is approximately proportional to $1/n_a$.
\bibitem{noteaboutPZW}
The Hamiltonian of Eq.~(\ref{eq:HPZW}) is a natural generalization of the familiar length gauge Hamiltonian generally used for calculating BBR shift, which is an advantage of this formulation.
\end{thebibliography}
\end{document}